\documentclass[lettersize,journal]{IEEEtran}
\newcommand{\Rmnum}[1]{\expandafter\@slowromancap\romannumeral #1@}
\usepackage{multicol}
\usepackage[utf8]{inputenc}
\usepackage[T1]{fontenc}
\usepackage{cite}
\usepackage{amsmath,amssymb,amsfonts}
\usepackage{algorithmic}
\usepackage{graphicx}
\usepackage{textcomp}
\usepackage{array}
\usepackage{url}
\usepackage{xcolor}
\usepackage{bm}
\usepackage{tabularx}
\usepackage{newunicodechar}
\newunicodechar{−}{-}
\usepackage{siunitx}
\usepackage{multirow}
\usepackage{bbding}
\usepackage{colortbl} 
\usepackage{tikz}
\usepackage{soul, color, xcolor}
\usetikzlibrary{positioning}
\usepackage[caption=false,font=normalsize,labelfont=sf,textfont=sf]{subfig}
\begin{document}

\title{
EEG-Based Emergency Braking Intensity Prediction Using Blind Source Separation}
\newcommand{\plainup}[1]{\textsuperscript{\normalfont#1}}  % 标准正体上标
\author{Zikun Zhou, Wenshuo Wang,~\IEEEmembership{Member,~IEEE}, Wenzhuo Liu,~\IEEEmembership{Graduate Student Member,~IEEE}, Hui Yao, Chaopeng Zhang,~\IEEEmembership{Student Member,~IEEE}, Yichen Liu, Xiaonan Yang, Junqiang Xi
\thanks{This work was supported by the National Natural Science Foundation of China (Grant No.:52572469, 52272411). The corresponding authors are \textit{Wenshuo Wang} and \textit{Junqiang Xi}.

All authors are with the School of Mechanical Engineering, Beijing Institute of Technology, Beijing, China {(email: zikun.zhou@bit.edu.cn; ws.wang@bit.edu.cn; wzliu@bit.edu.cn; hui.yao@bit.edu.cn; chaopeng666@gmail.com; yc.liu@bit.edu.cn; yangxn@bit.edu.cn; xijunqiang@bit.edu.cn.)}}
}

\maketitle

\begin{abstract}
Electroencephalography (EEG) signals have been promising for long-term braking intensity prediction but are prone to various artifacts that limit their reliability. Here, we propose a novel framework that models EEG signals as mixtures of independent blind sources and identifies those strongly correlated with braking action. Our method employs independent component analysis to decompose EEG into different components and combines time-frequency analysis with Pearson correlations to select braking-related components. Furthermore, we utilize hierarchical clustering to group braking-related components into two clusters, each characterized by a distinct spatial pattern. Additionally, these components exhibit trial-invariant temporal patterns and demonstrate stable and common neural signatures of the emergency braking process. Using power features from these components and historical braking data, we predict braking intensity at a $200$ ms horizon. Evaluations on the open source dataset (O.D.) and human-in-the-loop simulation (H.S.) show that our method outperforms state-of-the-art approaches, achieving $\mathrm{RMSE}$ reductions of $8.0\%$ (O.D.) and $23.8\%$ (H.S.).
\end{abstract}
\begin{IEEEkeywords}
Braking intensity prediction, electroencephalography, independent component analysis.
\end{IEEEkeywords}

\section{Introduction}
\IEEEPARstart{R}{ear-end} crash is one of the most frequent types of traffic accidents \cite{wang2022effect}, typically caused by abrupt deceleration of a leading vehicle and the delayed reaction from the following driver \cite{lyu2020towards}. To address this, Emergency Braking Assist (EBA) systems enhance braking force preemptively to avoid imminent collisions \cite{gunjate2023systematic,kusano2012safety}. Since early activation can improve safety, accurate prediction of the driver's braking intention is critical for timely intervention. Electroencephalography (EEG) captures neurophysiological activity preceding voluntary actions \cite{muller2008machine}, offering a promising approach to anticipate braking intention \cite{teng2017eeg} and enable earlier EBA deployment. Current EEG-based methods primarily treat braking prediction as a binary classification problem, predicting \textit{whether} braking will occur within a short time window. These methods fall into two categories: (i) Traditional machine learning, such as regularized linear discriminant analysis (RLDA) \cite{teng2017eeg,ju2022recognition,wang2017eeg,kim2014detection}, which counters overfitting in high-dimensional EEG data; (ii) Deep learning, particularly convolutional neural networks (CNNs) \cite{liang2023eeg,mora2023simplified,lutes2024convolutional,wenzhuo2025cvpr,wenzhuo2026}, which tackle challenges like high dimensionality, low signal-to-noise ratio, inter-subject variability, and diverse spectral information. Simplified CNNs have achieved over $84\%$ accuracy with only four electrodes \cite{mora2023simplified}, while hybrid CNN-spiking neural networks have achieved $99.06\%$ accuracy, underscoring the superiority of deep learning in this field \cite{lutes2024convolutional}.

\begin{figure}
    \centering
    \includegraphics[width=0.45\textwidth]{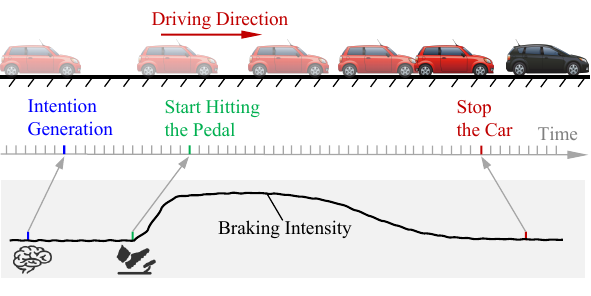}
    \caption{Illustration of an emergency braking scenario unfolding over time. The time interval between intention and action enables EEG-based braking intensity prediction, contributing to the construction of an adaptive control strategy for EBA. }
    \label{introduction}
\end{figure}

Despite these advances, existing methods fail to predict the braking intensity, a critical factor for safety-comfort tradeoffs in EBA systems. Fig. \ref{introduction} illustrates a typical emergency braking scenario, where the leading vehicle abruptly decelerates, prompting the following driver to brake and avoid a collision. If the predicted braking is insufficient, the EBA system supplements force to prevent a collision; if adequate, it avoids intervention to preserve comfort. Current approaches fail to meet these demands. Kim et al. \cite{kim2014decision} used kernel ridge regression (KRR) with multiple EEG features (e.g., event-related potentials, readiness potentials, and event-related desynchronization) but reported high prediction errors due to two limitations: (i) KRR's limited capacity to capture high-dimensional, nonlinear features, and (ii) contamination by artifacts (e.g., ocular, muscular), which degrades prediction performance \cite{zhang2025generative,li2025assessing}. We address the first limitation by employing deep neural networks which effectively model complex, high-dimensional nonlinear relationships in EEG data. To overcome the second, we apply blind source separation (BSS), which recovers underlying sources from mixed EEG observations, to raw EEG data to isolate sources and suppress artifacts \cite{bss}. However, some isolated sources weakly correlate with braking action, degrading prediction performance. To overcome this, we develop a multi-stage framework to systematically identify and retain braking-related sources. First, independent component analysis (ICA), a well-established BSS method, decomposes raw EEG signals into multiple independent components (ICs). We then evaluate the time-frequency patterns and Pearson correlations of ICs with braking intensity to select braking-related ICs. These ICs and historical braking data are fed into a deep learning model. Hierarchical clustering and time-domain analysis reveal distinct spatial topographies and temporal dynamics of braking-related ICs, suggesting stable and common neural substrates underlying braking behavior. Evaluations on open source dataset and human-in-the-loop simulation confirm our framework's superiority over baselines.

The contributions of this paper are twofold:

\begin{itemize}
\item An EEG-based framework for emergency braking intensity prediction. By isolating braking-related sources, our method effectively mitigates the influence of artifacts in EEG signals.
\item Identification of two distinct spatial patterns and trial-invariant temporal dynamics in braking-related independent components, exhibiting stable and common neural substrates underlying emergency braking behavior.
\end{itemize}

The remainder of this study is organized as follows. Section \uppercase\expandafter{\romannumeral2} introduces the framework of emergency braking intensity prediction. Section \uppercase\expandafter{\romannumeral3} describes the EEG datasets we utilized and corresponding feature extraction processes. Section \uppercase\expandafter{\romannumeral4} analyzes the spatial and temporal characteristics of braking-related ICs. Section \uppercase\expandafter{\romannumeral5} reports the results of comparative and ablation experiments. The conclusion is shown in Section \uppercase\expandafter{\romannumeral6}.
    
\section{The Framework of Emergency Braking Intensity Prediction}
\begin{figure}
\centerline{\includegraphics[width=0.5\textwidth]{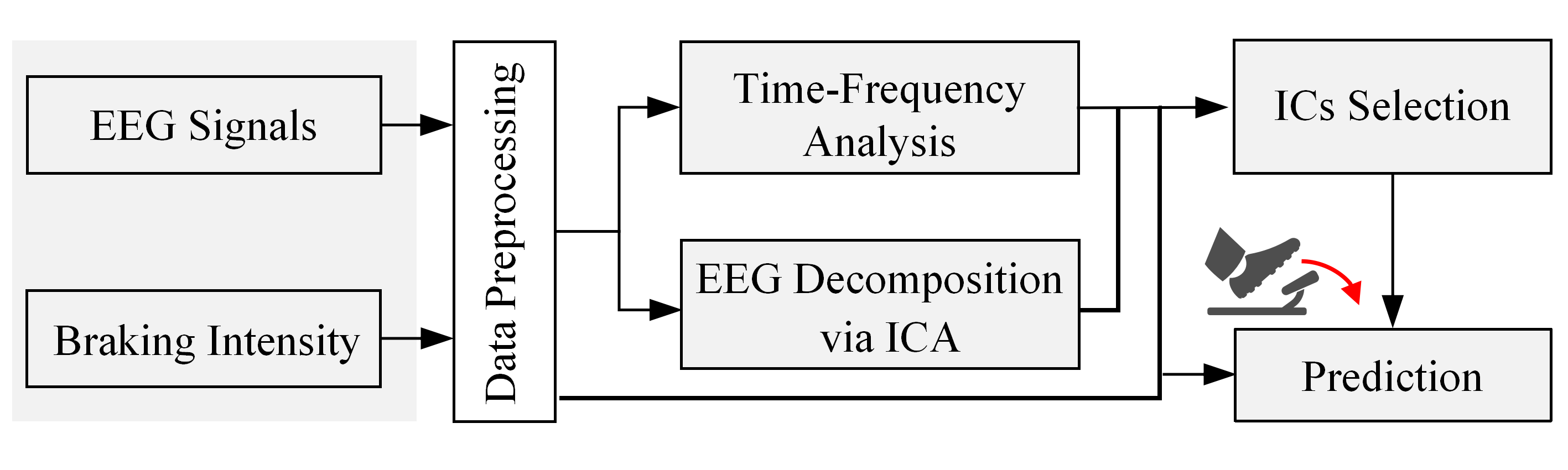}}
\caption{The framework of our proposed method for emergency braking intensity prediction with EEG.}
\label{figstart}
\end{figure}

Our proposed framework of predicting emergency braking intensity using EEG signals is shown in Fig. \ref{figstart}, which comprises four components: (i) Time–Frequency Analysis, which identifies EEG frequency bands associated with braking intention; (ii) EEG Decomposition via ICA, separating preprocessed EEG into ICs; (iii) ICs Selection, identifying braking-related ICs based on their correlations with braking action; and (iv) Prediction, integrating selected ICs and historical braking data to predict braking intensity. First, EEG and braking data are preprocessed for noise removal. The cleaned EEG signals undergo time-frequency analysis to identify frequency bands strongly linked to braking intention. Concurrently, ICA decomposes the EEG signals into ICs, revealing latent sources. Spectral power within the identified frequency band is computed as IC feature. We then use Pearson correlation analysis to evaluate the relationship between these features and braking intensity, retaining only ICs with significant correlations to ensure biological relevance and reduce dimensionality. Finally, spectral features from braking-related ICs and historical braking intensity data are fed into a multilayer perceptron (MLP) model for prediction.

\subsection{Time-Frequency Analysis}

EEG signals exhibit oscillations across distinct frequency bands \cite{luck2014introduction}, whose amplitudes are quantified as spectral power, a feature linked to brain states and cognitive functions \cite{stevens2019creativity,ouyang2020decomposing}. This justifies its use for decoding braking intention. Traditional power analysis based on Fourier Transform lacks sufficient temporal resolution \cite{morales2022time}; therefore, we adopt time-frequency analysis to capture dynamic spectral power changes. In this paper, we apply Event-Related Spectral Perturbation (ERSP) to visualize spectral power variations during braking \cite{vecchio2022time}.

\subsection{EEG Decomposition via ICA}
\subsubsection{ICA}
The neuronal electrical activity underlying EEG is primarily generated by postsynaptic potentials, whose prolonged duration allows summation across populations of aligned, similarly oriented pyramidal neurons \cite{luck2014introduction}. The activity of such synchronously active local population can be modeled as an equivalent current dipole, corresponding to a distinct neural source. Its time-varying amplitude is termed \textit{source waveform} or \textit{source signal}. 

\begin{figure}[t]
\centerline{\includegraphics[width=0.5\textwidth]{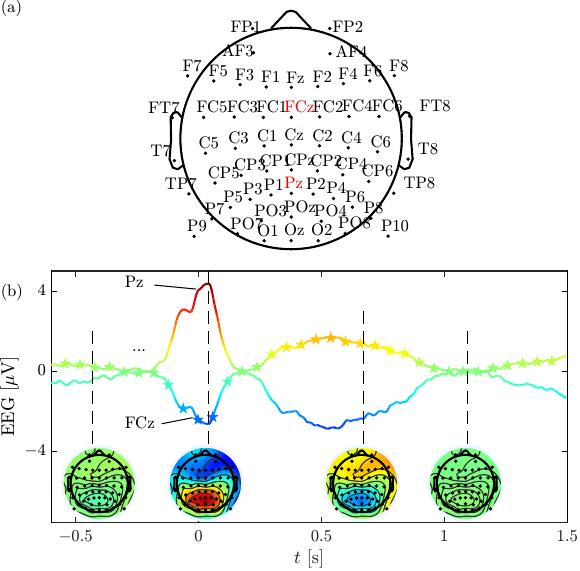}}
\caption{The illustrations of (a) electrode location with FCz and Pz highlighted in red and (b) an IC with its scalp map evolving over time.}
\label{fig4}
\end{figure}

Consider $m$ neural sources in the brain, each characterized by a source waveform $\mathbf{s}_i = [s_{i,1}, \dots, s_{i,n}]\in \mathbb{R}^n$ representing its activity over $n$ time points. Assuming the number of sources equals the number of electrodes ($m$), the EEG records at electrode $i$ is $\mathbf{x}_i = [x_{i,1}, \dots, x_{i,n}]^{\top}\in \mathbb{R}^n$. The EEG data matrix $\mathbf{X}$ is constructed as $\mathbf{X}=[\mathbf{x}_1,\dots,\mathbf{x}_m]^{\top}\in\mathbb{R}^{m \times n}$, with each row corresponding to one electrode's time series. Due to volume conduction, the EEG signal at electrode $i$ is modeled as a linear mixture of source waveforms:

\begin{equation}
    \mathbf{x}_{i} = a_{i1}\mathbf{s}_{1}^{\top} + \dots + a_{ij}\mathbf{s}_{j}^{\top} + \dots + a_{im}\mathbf{s}_{m} ^{\top}
    \label{xasas}
\end{equation}
where $a_{ij}\in\mathbb{R}$ is the projection weight of the $j$-th source to the $i$-th electrode, determined by spatial geometry and tissue conductivity \cite{makeig2011erp}. The EEG signals matrix $\mathbf{X}$ is decomposed as: 

\begin{equation}
\begin{aligned}
\mathbf{X} &=\sum_{i=1}^m \mathbf{a}_i\mathbf{s}_{i} = 
\begin{bmatrix}
    a_{11} \\ 
    a_{21} \\
    \vdots \\
    a_{m1}
\end{bmatrix}
\mathbf{s}_{1}
+
\dots
+
\begin{bmatrix}
    a_{1m} \\ 
    a_{2m} \\
    \vdots \\
    a_{mm}
\end{bmatrix}
\mathbf{s}_{m}
\label{XAS1}
\end{aligned}
\end{equation}
where $\mathbf{a}_{i}=[a_{1i},\dots,a_{mi}]^{\top}$ maps the $i$-th sources $\mathbf{s}_i$ to $\mathbf{X}$. Each outer product $\mathbf{a}_{i}\mathbf{s}_i$ corresponds to an independent component. The IC's scalp map, i.e., its spatial potential distribution, is determined by $\mathbf{a}_{i}$ and scaled by $\mathbf{s}_i$. For instance, at time $j$, the scalp map is $\mathbf{a}_{i}s_{i,j}$. 
Fig. \ref{fig4} visualizes the time evolution of an IC's scalp map, with red and blue indicating positive and negative potentials, respectively, and color intensity reflecting absolute strength. The scalp map exhibits a consistent spatial pattern across time, with amplitudes linearly scaled by $\mathbf{s}_{i}$. Electrode potentials (e.g., FCz, Pz) share the same temporal waveform $\mathbf{s}_{i}$, scaled by $\mathbf{a}_{i}$'s entries.
The source matrix $\mathbf{S}$ is recovered via linear combinations of $\mathbf{X}$:  
\begin{equation}
\begin{aligned}
\mathbf{S} =\mathbf{W}\mathbf{X}\label{SWX}
=
\begin{bmatrix}
    \mathbf{w}_1 \\ 
    \mathbf{w}_2 \\
    \vdots \\
    \mathbf{w}_m
\end{bmatrix}
\mathbf{X}
=
\begin{bmatrix}
    \mathbf{s}_1 \\
    \mathbf{s}_2 \\
    \vdots \\
    \mathbf{s}_m
\end{bmatrix}
\end{aligned}
\end{equation}
where each row $\mathbf{w}_i$ of $\mathbf{W}=\mathbf{A}^{-1}$ acts a spatial filter extracting the $i$-th source waveform $\mathbf{s}_i=\mathbf{w}_i\mathbf{X}$. 

\subsubsection{Model Development of ICA}
ICA assumes that the source signals are statistically independent, meaning that the value of one signal neither influences nor predicts another. This independence can be quantified by the mutual information $\mathrm{I}(s_1,s_2)$ (take two sources as an example) \cite{kraskov2004estimating}, defined as:

\begin{equation}
\mathrm{I}(s_1,s_2)
=
\mathrm{H}(s_1)-\mathrm{H}(s_1|s_2)
\label{hxx}
\end{equation}
where $\mathrm{H}(s_1)$ is the entropy of $s_1$, and $\mathrm{H}(s_1|s_2)$ is the conditional entropy of $s_1$ given $s_2$. 

\textbf{Mutual Information Minimization and Joint Entropy Maximization.} To separate sources from mixed EEG signals, ICA minimizes the mutual information among the estimated sources. Alternatively, this problem can be formulated as maximizing their joint entropy \cite{bell1995information}. Given two sources $s_1$ and $s_2$, the joint entropy is:

\begin{equation}
\mathrm{H}(s_{1},s_{2})
=
\mathrm{H}(s_{1})
+
\mathrm{H}(s_{2})
-\mathrm{I}(s_{1},s_{2})
\label{joint_entropy}
\end{equation}
In practice, the interferences of $\mathrm{H}(s_{1})$ and $\mathrm{H}(s_{2})$ to $\mathrm{I}(s_{1},s_{2})$ are often negligible \cite{bell1995information}. Therefore, minimizing mutual information is equivalent to maximizing $\mathrm{H}(s_{1},s_{2})$.

\textbf{Entropy Formulation for Single Sources.} For a single source waveform, let $s,x$ denote scalar random variables representing the source and EEG amplitude at an arbitrary time point, respectively. The entropy $\mathrm{H}(s)$ is defined by its probability density function (PDF) $f_{s}(s)$:

\begin{equation}
    \mathrm{H}(s) = -\mathbb{E} \left[ \ln f_{s}(s) \right] = -\int_{-\infty}^\infty f_{s}(s) \ln f_{s}(s) \, ds
    \label{jifen}
\end{equation}
where $\mathbb{E}[\cdot]$ denotes expectation. If $s$ is derived from $x$ via a monotonic transformation, the PDF $f_{s}(s)$ is given by \cite{papoulis1965random}:

\begin{equation}
   f_{s}(s)= \frac{f_{x}(x)}{\left|\frac{\partial s}{\partial x}\right|},
   \label{eq:pdf_transformation}
\end{equation}
Substituting Eq. \eqref{eq:pdf_transformation} into Eq. \eqref{jifen} yields:

\begin{equation}
    \mathrm{H}(s) = \mathbb{E}\left[\ln\left|\frac{\partial s}{\partial x}\right|\right] - \mathbb{E}\left[\ln f_{x}(x)\right]
\end{equation}
In practice, a nonlinear invertible transformation is used instead of a linear one to capture higher-order moments of the PDF:

\begin{equation}
    s = g(wx+b)
    \label{nonlinearfunction}
\end{equation}
where $w$ and $b$ are the weight and bias parameters, respectively, and $g$ is an invertible nonlinear function (e.g., hyperbolic tangent). To maximize $\mathrm{H}(s)$, the parameters $w$ and $b$ can be optimized using gradient-based methods.

\textbf{Extension to Multiple Sources.} Let $\mathbf{s}=[s_1,\dots,s_m]\in\mathbb{R}^{m}$ denote the vector of source amplitudes at an arbitrary time point. The multivariate PDF under an element-wise monotonic transformation $\mathbf{s} = g(\mathbf{W}\mathbf{x}+\mathbf{b})$ is:

\begin{equation}
    f_{\mathbf{s}}(\mathbf{s})=\frac{f_{\mathbf{x}}(\mathbf{x})}{\left|\det(\mathbf{J})\right|}
\end{equation}
where $\left|\det(\mathbf{J})\right|$ is the absolute value of the determinant of the partial derivatives matrix $\mathbf{J}$:

\begin{equation}
    \mathbf{J} = \begin{bmatrix}
\frac{\partial s_1}{\partial x_1} & \dots & \frac{\partial s_1}{\partial x_m} \\
\vdots & & \vdots \\
\frac{\partial s_m}{\partial x_1} & \dots & \frac{\partial s_m}{\partial x_m}
\end{bmatrix}
\end{equation}
The joint entropy $\mathrm{H}(\mathbf{s})$ is then given by:

\begin{equation}
    \mathrm{H}(\mathbf{s}) = \mathbb{E}\left[\ln\left|\det(\mathbf{J})\right|\right] - \mathbb{E}\left[\ln f_{\mathbf{x}}(\mathbf{x})\right]
    \label{HS}
\end{equation}
The partial derivatives of $\mathrm{H}(\mathbf{s})$ with respect to $\mathbf{W}$ and $\mathbf{b}$ are computed via the chain rule, enabling optimization via gradient ascent.

\subsection{ICs Selection}
\begin{figure}[t]
\centerline{\includegraphics[width=0.5\textwidth]{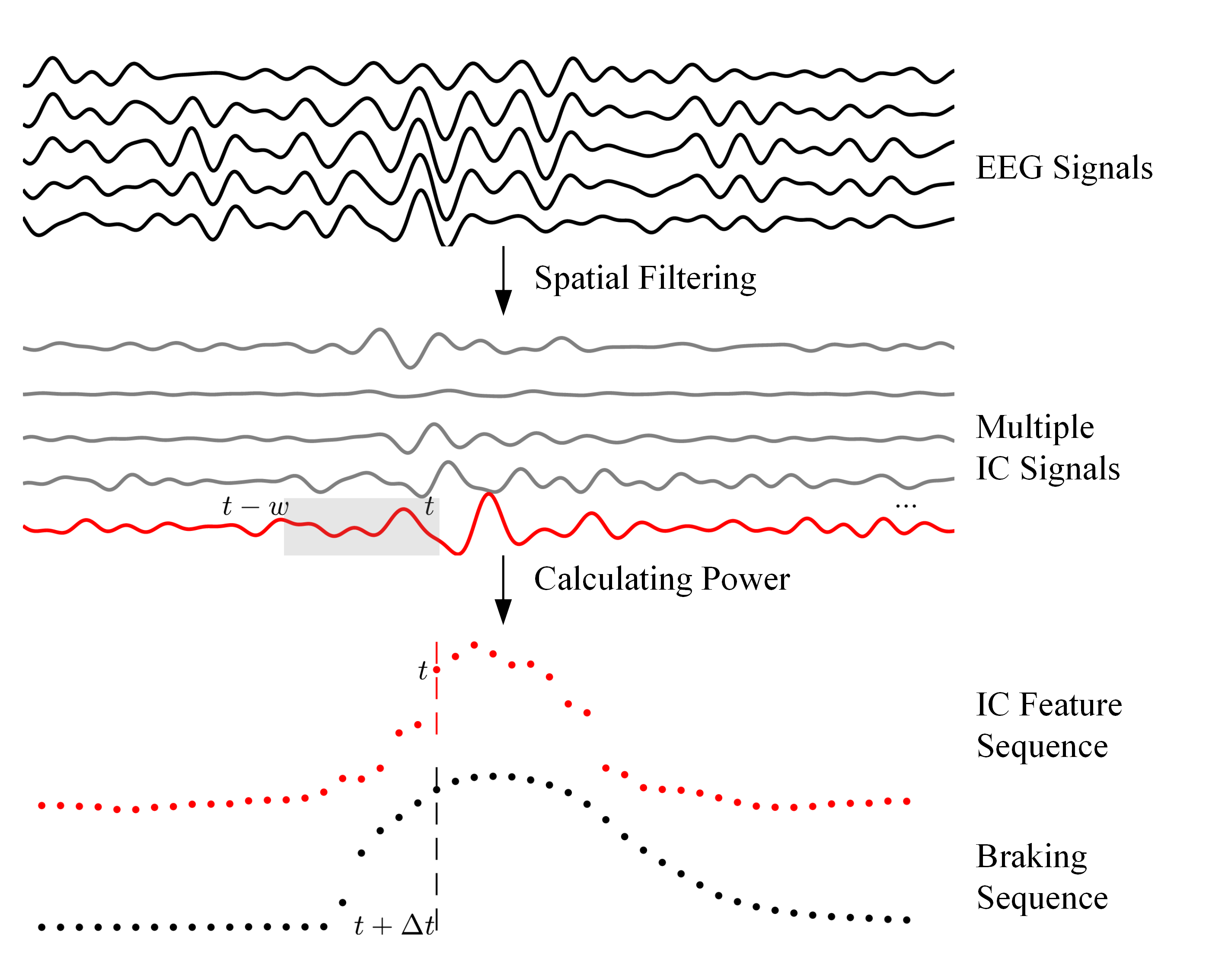}}
\caption{The procedure for calculating the Pearson correlation coefficient between a pair of IC feature sequence and braking sequence.}
\label{pcc}
\end{figure}

Preprocessed EEG signals were decomposed into ICs, each corresponding to a distinct physiological process. To identify braking-related ICs, we evaluated their relationship with braking intensity using Pearson correlation coefficients. Fig. \ref{pcc} shows the analysis procedure. First, for each IC, the power of its source signal within the time-frequency-identified band was computed as the IC feature (see Appendix A). Using a sliding window of length $w$ (see Appendix B), the source signal was divided into epochs ranging from $t-w$ to $t$. For each time point $t$, the IC feature was paired with braking intensity sampled at $t+\Delta t$, where $\Delta t\in\{200,300,400\}$ ms to account for neural delays. For each trial, the Pearson correlation coefficient was then computed between the IC feature sequence and the delayed braking sequence. A coefficient exceeding $0.6$ indicated a strong positive correlation within the trial. To quantify the overall association strength, we defined the metric:

\begin{equation}
\nu=\frac{N_{\mathrm{strong}}}{N_{\mathrm{total}}},\label{ratio}
\end{equation}
where $N_{\mathrm{strong}}$ is the number of trials with strong positive correlations, and $N_{\mathrm{total}}$ is the total braking trials. High $\nu$ values reflect strong and consistent IC-braking relationships across trials.

For each subject, we ranked the ICs in descending order based on their $\nu$ values to obtain an initial set of candidate braking-related ICs. However, these candidates often included artifacts such as ocular movements, electromyographic signals, and electrode noise \cite{bhattacharyya2022ocular}, which could impair subsequent analysis. To address this, we used ICLabel for artifact rejection. ICLabel classifies ICs into seven categories: six artifacts (`ocular', `muscular', `cardiac', `line noise', `channel noise', and `other') and neural components `brain' \cite{pion2019iclabel}. We excluded ICs with highest-probability assigned to any of the five artifact categories (`ocular', `muscular', `cardiac', `line noise', and `channel noise'), as they were definitively non-neural. ICs labeled as `other' were retained despite potential artifacts, as they might still contain neural components. From the remaining ICs, we selected the top $5\%$ with the highest $\nu$ values as braking-related ICs.

\subsection{Prediction}

We employed a Multi-layer Perceptron (MLP) due to its proven effectiveness in EEG-based motion trajectory prediction \cite{pancholi2022source}. The network architecture comprises the following components:
\begin{itemize}
    \item Input layer: Braking-related IC power $\mathbf{p}_{t}^{\mathrm{ic}}$ and braking intensity $\mathbf{y}_{t}^{\mathrm{his}}$, both sampled over historical time steps.
    \item Hidden layers: Four fully connected layers with $30$, $20$, $10$, and $5$ neurons.
    \item Activation function: Parametric rectified linear unit (PReLU) \cite{he2015delving}, which generalizes ReLU by learning negative slopes to alleviate the dying ReLU problem.
    \item Output layer: Predicted braking intensity at time $t+\Delta t$:

    \begin{equation}
    y_{t+\Delta t}^{\mathrm{pre}}=\mathrm{MLP}(\mathbf{p}^{\mathrm{ic}},\mathbf{y}^{\mathrm{his}})
    \label{mlp}
\end{equation}
where $\mathbf{p}^{\mathrm{ic}}=[p_{t-(m-1)h:t}^{\mathrm{ic_1}},\dots,p_{t-(m-1)h:t}^{\mathrm{ic_r}}]$, $r$ is the number of braking-related ICs, $h = 50$ ms, $m = 5$, and $\mathbf{y}^{\mathrm{his}}=y_{t-(m-1)h:t}^{\mathrm{his}}$. The prediction horizon $\Delta t$ is selected from $\{200,300,400\}$ ms (see Section \uppercase\expandafter{\romannumeral3}-D). 
\end{itemize}

Here, we performed prediction with a step size of $50$ ms. During training, we minimized the mean square error (MSE) loss using the Adam optimizer \cite{kingma2014adam}.

\section{Data Description and Processing}
This section describes the two datasets employed, including their experimental protocols and recording setups. Since their feature extraction processes are similar, we detail the procedure only for the open source dataset.

\subsection{Data Description}
\begin{figure}
\centerline{\includegraphics[width=0.5\textwidth]{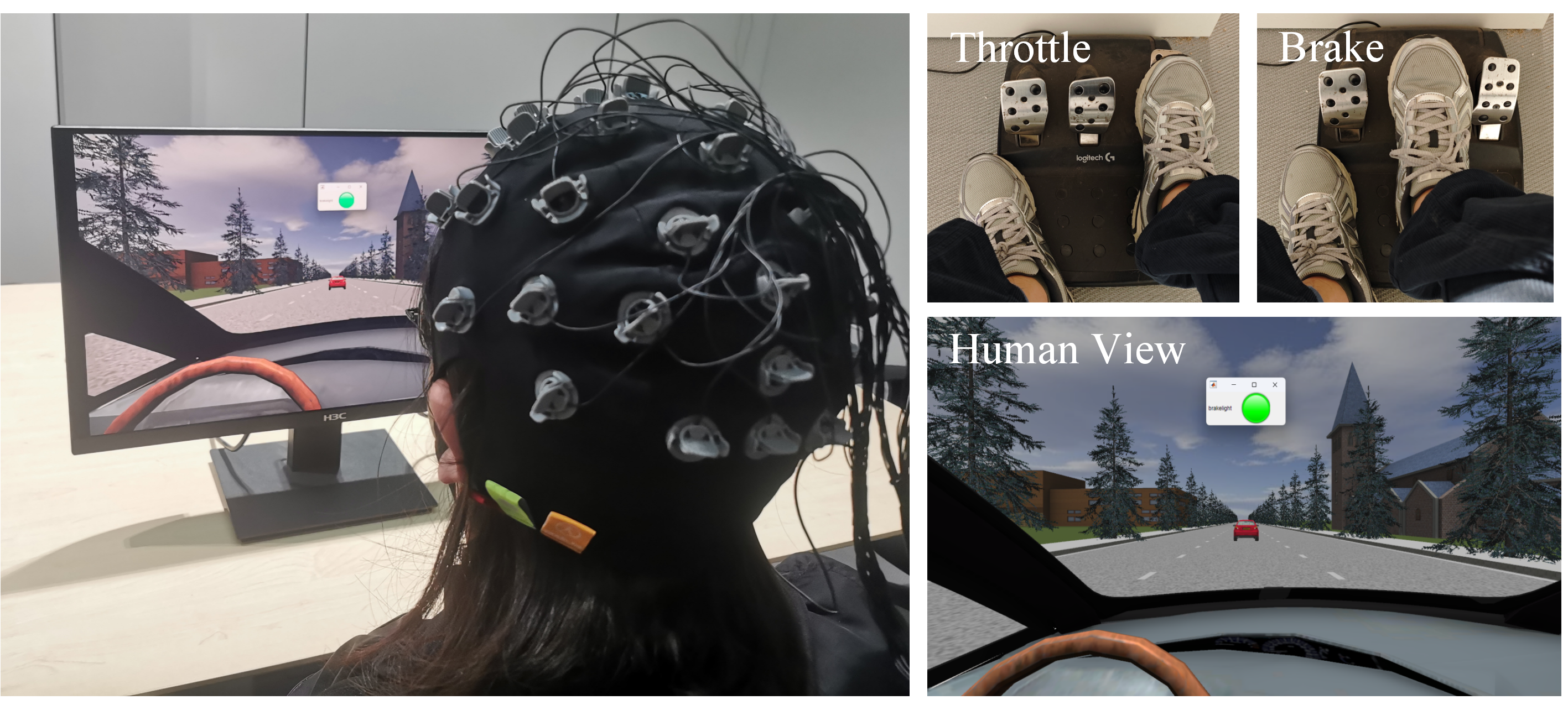}}
\caption{The experimental setup for emergency braking.}
\label{online}
\end{figure}
\subsubsection{Open Source Dataset (O.D.)}
The dataset comprises EEG signals from a driving simulation experiment by Haufe et al. \cite{haufe2011eeg}. Eighteen participants controlled a virtual vehicle through a steering wheel, accelerator, and brake pedal, maintaining a safe distance behind a computer-controlled lead vehicle. If the inter-vehicle gap decreased below $20$ m, the lead vehicle abruptly decelerated with activated brake lights, prompting participants to execute emergency braking maneuvers. Braking intensity signals were acquired at $67$ Hz, and EEG signals were recorded at $1000$ Hz via a $59$-channel cap. Both signals were resampled to $200$ Hz for synchronization.

\subsubsection{Human-in-the-Loop Simulation (H.S.)}
We conducted an emergency braking experiment with $14$ participants ($9$ males and $5$ females) for data collection, with a mean age of $23.3$ years (SD $=1.7$). All participants provided written informed consent and were informed about their right to withdraw at any time without consequence. As shown in Fig. \ref{online}, participants seated facing a LED display presenting driving simulation environment created with Prescan/Simulink. Their task was to control a ego vehicle closely following a computer-controlled red lead vehicle on a $35$-km straight road. When the headway distance dropped below $30$ m, the lead vehicle randomly decelerated from $90$ km/h to $40$ km/h at $10$-$15$ s, accompanied by the brake light switching from green to red. Participants were instructed to brake emergently to avoid collisions. Brake pedal deflections were recorded at $20$ Hz, and EEG signals were acquired at $256$ Hz via a $32$-channel cap. Both signals were resampled to $200$ Hz for synchronization. 

\subsection{Data Preprocessing}
Following the established preprocessing protocols \cite{gao2025cross,hu2024eeg,palaniappan2021investigating,lyu2025driver}, we processed the raw EEG data as follows (parameters are detailed in Appendix B). First, a $45$ Hz low-pass filter was applied to remove power line interference ($50$ Hz). Second, EEG data were segmented into epochs spanning $\pm1500$ ms relative to brake pedal activation onset, capturing the short-duration braking events ($\sim 3$ s) while excluding long inter-trial intervals ($>10$ s). Third, each epoch was baseline-corrected by subtracting the mean signal amplitude during a pre-stimulus window to mitigate instrumental drift and motion artifacts. Finally, a common average reference was performed by subtracting the global mean across all channels at each time point to reduce noise induced by the reference electrode.

\begin{figure}[t]
\centerline{\includegraphics[width=0.45\textwidth]{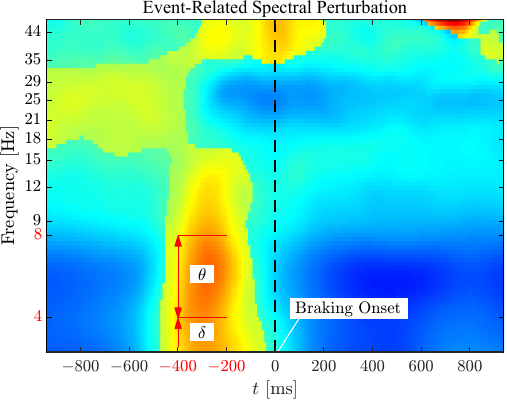}}
\caption{The illustration of grand-average ERSP across all participants during emergency braking. Spectral power is depicted using a color gradient, with blue indicating suppression and red indicating enhancement relative to baseline.}
\label{fig5}
\end{figure}

\subsection{Time-Frequency Analysis}
\label{AA}

Fig. \ref{fig5} shows the grand-average ERSP across all participants during emergency braking. The time window spans from $-800$ ms to $800$ ms and is aligned to the braking onset (black dashed line at $t=0$ ms). The frequency band ($3$-$45$ Hz) includes the $\theta$ ($4$-$8$ Hz), $\alpha$ ($8$-$12$ Hz), and $\beta$ ($12$-$30$ Hz) bands, partially covering the $\delta$ ($0.5$-$4$ Hz) and $\gamma$ ($30$-$100$ Hz) bands. Spectral power is depicted using a color gradient, with blue indicating suppression and red indicating enhancement relative to baseline. A pronounced power increase (red) in the $\delta$-$\theta$ band emerges at $200$-$400$ ms before braking onset, followed by suppression (blue) post-onset. This temporal-spectral pattern suggests a functional link between low-frequency oscillations and braking intention, prompting our selection of $\delta$ and $\theta$ power as discriminative features for analysis.

\subsection{Braking-Related IC selection}

\begin{figure}[t]
\centerline{\includegraphics[width=0.47\textwidth]{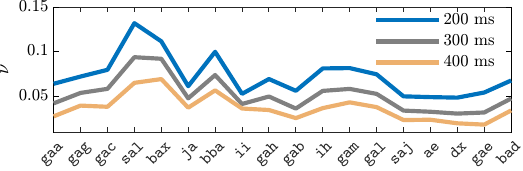}}
\caption{Illustration of average correlation $\nu$ across all subjects with the prediction horizon of $200$, $300$, $400$ ms, respectively.}
\label{fig6}
\end{figure}

Fig. \ref{fig6} shows the average correlation metric $\nu$ across all subjects for prediction horizons of $200$ ms, $300$ ms, $400$ ms. The results reveal an inverse relationship between prediction horizon and correlation strength: the $200$ ms horizon (blue curve) exhibiting the highest $\nu$ consistently across all subjects. Therefore, we selected $200$ ms as the optimal prediction horizon for analysis.

\begin{figure}
\centerline{\includegraphics[width=0.47\textwidth]{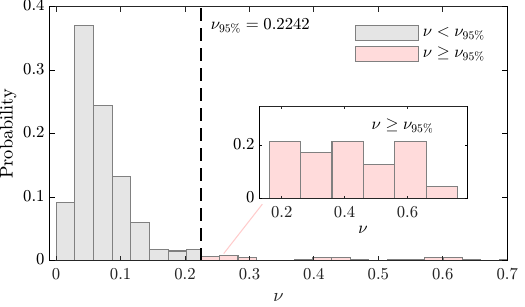}}
\caption{Histogram of $\nu$ values for all retained ICs, alongside the distribution of the top $5\%$ ($\nu>\nu_{95}$).}
\label{nu_distribution}
\end{figure}

For the $200$ ms horizon, we removed artifacts using ICLabel and showed the distribution of $\nu$ values for the remaining ICs in Fig. \ref{nu_distribution}. The correlation metric $\nu$ ranges from $0$ to $0.713$ across all retained ICs, with the majority concentrated near $0.05$. ICs exceeding the $95$th percentile ($\nu_{95}=0.2242$) were classified as braking-related. The $\nu$ values of these braking-related ICs (top $5\%$) exhibited a relatively uniform distribution spanning $0.2242$ to $0.713$, distinct from the sharp peak observed in the full IC set.

\section{Analysis for Braking-Related ICs}
This section analyzes the spatial and temporal patterns of braking-related ICs extracted by our method. Hierarchical clustering of these ICs in O.D. reveals two dominant scalp map patterns across subjects. Moreover, in H.S., both average and individual EEG waveforms exhibit temporally consistent patterns. These results indicate stable and common neural substrates for braking intention generation.

\subsection{Analysis for Open Source Dataset}
\subsubsection{Hierarchical Clustering}
Based on Section \uppercase\expandafter{\romannumeral2}, we identified $23$ braking-related ICs across all subjects (excluding \texttt{gae}), each exhibiting distinct scalp maps. To determine whether these ICs share similar spatial patterns, we performed hierarchical clustering with scalp maps as input. 

\begin{figure}[t]
\centerline{\includegraphics[width=0.47\textwidth]{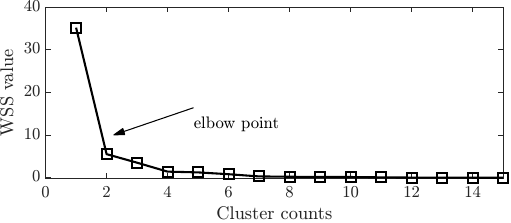}}
\caption{The WSS values for different numbers of clusters.}
\label{num}
\end{figure}

In EEGLAB, each IC's scalp map was represented as a $67\times67$ matrix \cite{delorme2004eeglab}, where scalp regions contained potential values and non-scalp regions were assigned `NaN'. To address polarity inconsistencies, we inverted maps with opposite polarity relative to their actual scalp maps. After excluding `NaN' values, each matrix was flattened into a $3409$-dimensional vector. Given its high dimensionality, we applied Principal Component Analysis (PCA) \cite{gewers2021principal} and retained principal components explaining $95\%$ of the cumulative variance. 

\begin{figure}
\centering
\includegraphics[width=0.47\textwidth]{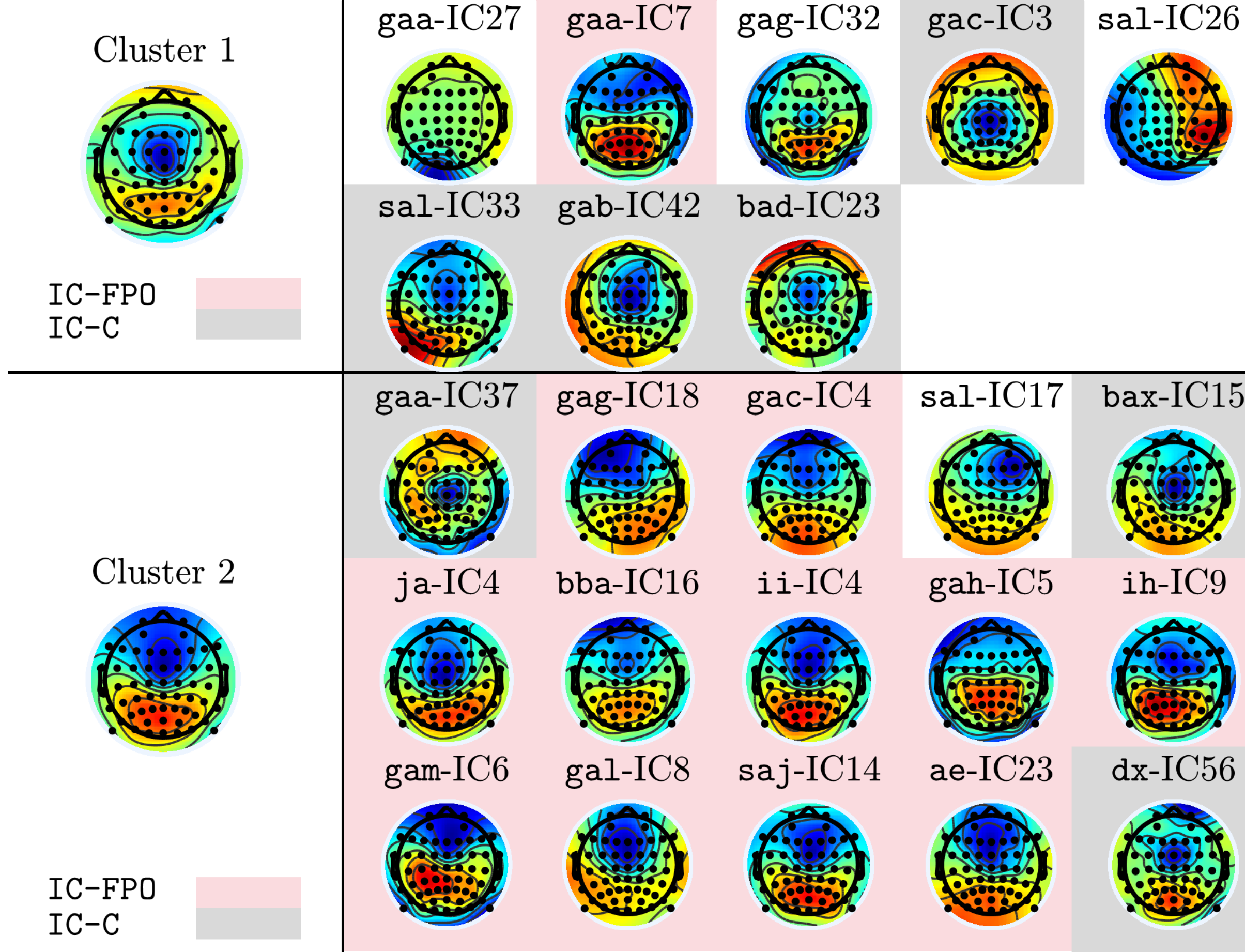}
\caption{Illustration of two clusters, alongside their centroids. Scalp maps consistent with the centroids of cluster $1$ and $2$ are highlighted in light-gray and light-pink regions, respectively.}
\label{fig:IC-FB-C}
\end{figure}

We employed the unweighted pair group method with arithmetic mean (UPGMA) \cite{sokal1958statistical}. The algorithm iteratively: compute a pairwise distance matrix, merge the two nearest clusters, and update the distance matrix using the UPGMA linkage criterion. We used cosine distance as the metric because it is scale-independent \cite{cos2022}. The UPGMA linkage criterion updates distances between merged and existing clusters via proportional averaging. For example, if clusters $\mathcal{A},\mathcal{B}$ with size $|\mathcal{A}|,|\mathcal{B}|$ are merged, the distance between $\mathcal{A}\cup\mathcal{B}$ and another cluster $\mathcal{C}$ is:

\begin{equation}
    d_{(\mathcal{A}\cup\mathcal{B}),\mathcal{C}} = \frac{|\mathcal{A}| \cdot d_{\mathcal{A},\mathcal{C}} + |\mathcal{B}| \cdot d_{\mathcal{B},\mathcal{C}}}{|\mathcal{A}| + |\mathcal{B}|}
    \label{update}
\end{equation}
where the updated distance $d_{(\mathcal{A}\cup\mathcal{B}),\mathcal{C}}$ is derived by the proportional averaging of $d_{\mathcal{A},\mathcal{C}}$ and $d_{\mathcal{B},\mathcal{C}}$.

To determine the optimal number of clusters $K$, we applied the elbow method by computing the Within-Cluster Sum of Squares (WSS) for $K\in [1,15]$. As shown in Fig, \ref{num}, a noticeable inflection point occurs at $2$, beyond which the marginal reduction in WSS becomes negligible, indicating that the data naturally partition into two clusters.

\begin{figure}[t]
    \centering
    \includegraphics[width=0.48\textwidth]{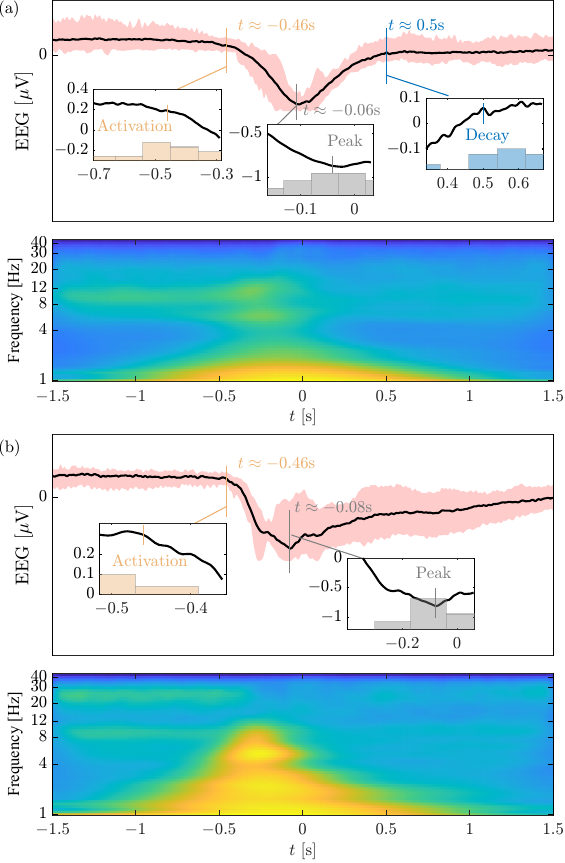}
    \caption{Illustration of grand-average potentials for cluster (a) \texttt{IC-FPO} and (b) \texttt{IC-C} with corresponding spectrograms.}
    \label{FPO_TF}
\end{figure}

\subsubsection{Interpretation of Hierarchical Clustering}
  
Hierarchical clustering revealed two clusters with distinct spatial patterns (Fig. \ref{fig:IC-FB-C}). Scalp maps showing consistency with the centroids of cluster $1$ and $2$ are highlighted in light-gray and light-pink regions, respectively.

The pink-highlighted ICs demonstrated a stable configuration: negative potential in frontal regions coupled with positive potentials in parietal-occipital regions. We designated these ICs as \texttt{IC-FPO} (frontal–parietal–occipital). Their bipolar distribution suggests two functional roles: (i) Parietal/occipital contributions, integrating visual and vestibular inputs for spatial navigation \cite{leon1998end,thier1997parietal}, and (ii) frontal contributions, supporting executive functions such as attention, decision-making, and planning \cite{schall2002monitoring,praamstra2005frontoparietal}.

Fig. \ref{FPO_TF}(a) presents the grand-average potentials and spectrogram of \texttt{IC-FPO}, including the \textcolor[rgb]{0.93,0.69,0.43}{activation}, \textcolor[rgb]{0.5,0.5,0.5}{peak}, and \textcolor[rgb]{0,0.4470,0.7410}{decay} to baseline timings across all ICs. The red-shaded region represents the inter-subject variability range of potentials across all ICs, while the black trace depicts the mean potential. A negative deflection begins around $-460$ ms (relative to braking onset, $t=0$ ms), marking the activation of \texttt{IC-FPO}. At the individual-subject level: (i) \textcolor[rgb]{0.93,0.69,0.43}{Activation} onsets occur between $-700$ ms to $-300$ ms, (ii) \textcolor[rgb]{0.5,0.5,0.5}{Peak} onsets occur predominantly between $-100$ ms and $0$ ms, with occasional post-braking onset delays, and (iii) \textcolor[rgb]{0,0.4470,0.7410}{Decay} to baseline onsets span from $340$ ms to $670$ ms, aligning with the grand-average baseline return ($\approx 500$ ms). The spectrogram shows prolonged $\delta$-band ($1$-$2$ Hz) power enhancement ($-1000$ ms to $1000$ ms), with greater duration and higher power compared to the transient $\theta$-band (from $-500$ ms to $0$ ms) response.

The gray-highlighted background exhibited localized negative potentials, distinct from surrounding regions. These ICs were classified as \texttt{IC-C} (Central). The central region corresponds to the primary motor cortex, suggesting involvement in lower-limb motor control (e.g., knee and ankle joint movements during braking) \cite{naito2004sensing}. Fig. \ref{FPO_TF}(b) shows \texttt{IC-C}'s grand-average waveform with a negative deflection starting at $-460$ ms and temporally synchronized with \texttt{IC-FPO}. At the individual-subject level, \textcolor[rgb]{0.93,0.69,0.43}{activation} occurs between $-500$ ms and $-400$ ms, indicating a similar initiation period. Additionally, \texttt{IC-C} \textcolor[rgb]{0.5,0.5,0.5}{peaks} slightly earlier ($\approx -80$ ms) and remains active until $1500$ ms, markedly longer ($1000$ ms) than \texttt{IC-FPO}. Spectral analysis further distinguishes \texttt{IC-C} from \texttt{IC-FPO}. \texttt{IC-C} shows similar $\delta$- and $\theta$-band power amplitudes, contrasting with the \texttt{IC-FPO}'s $\delta$-dominant profile. Moreover, \texttt{IC-C}'s $\delta$ ($1$-$4$ Hz) and $\theta$ ($4$-$8$ Hz) activity spans broader frequency ranges than \texttt{IC-FPO}.

\subsection{Analysis for Human-in-the-Loop Simulation}
\begin{figure}
    \centering
    \includegraphics[width=0.47\textwidth]{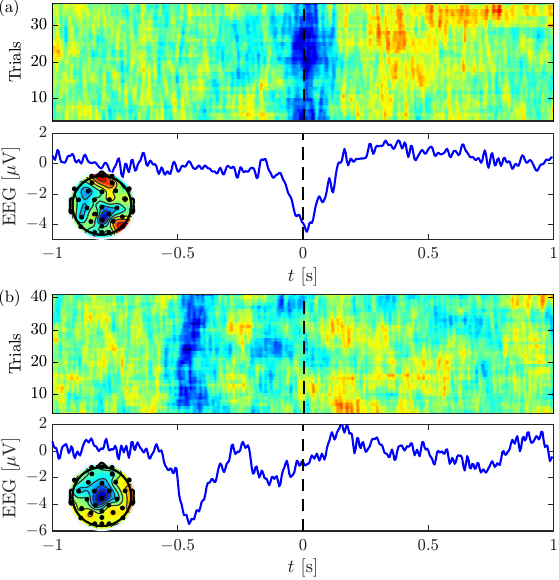}
    \caption{Illustration of the average and individual EEG waveforms of the braking-related IC in (a) the female and (b) the male driver, with their scalp maps.}
    \label{female}
\end{figure}

% In the H.S. dataset, while braking-related ICs do not exhibit consistent scalp maps like \texttt{IC-FPO} or \texttt{IC-C}, their temporal patterns are consistent across trials.
In the H.S. dataset, braking-related ICs do not exhibit consistent scalp maps like \texttt{IC-FPO} or \texttt{IC-C}, likely attributable to the fewer electrodes in H.S. ($32$ versus $59$ in O.D.) \cite{howmanyelectrodes}. Notably, their temporal patterns remain consistent across trials. For instance, Fig. \ref{female}(a) shows the average and individual EEG waveforms of a braking-related IC from one female driver, along with its non-canonical scalp map. All trials are aligned and visualized as a color map, with potentials displayed in blue (negative) or red (positive). The average waveform shows a negative deflection starting at $-100$ ms, peaking at braking onset, and returning toward baseline by $200$ ms. Most trials exhibit synchronized negative potentials around $-100$ ms, confirming the temporal reliability of this IC. 

Fig. \ref{female}(b) shows the average and individual EEG waveforms of the braking-related IC in one male driver, along with its scalp map, which exhibits a spatial pattern similar to \texttt{IC-C}. The average waveform displays a negative deflection around $-500$ ms. Most trials align temporally with the average waveform, as evidenced by a consistent blue-shaded region near $-500$ ms. These two braking-related ICs show a significant functional association with braking behavior. 

\renewcommand{\arraystretch}{1.25}
\begin{table*}[!t]
\caption{The Prediction Performance of Different Methods with the Input of Braking Intensity. O.D.: Open Source Dataset, H.S.: Human-in-the-loop Simulation. The Best Results are Marked in $\mathbf{BOLD}$.}
\label{table1}
\centering
\begin{tabular}{
>{\columncolor[HTML]{FFFFFF}}l 
>{\columncolor[HTML]{FFFFFF}}c |
>{\columncolor[HTML]{FFFFFF}}c 
>{\columncolor[HTML]{FFFFFF}}c 
>{\columncolor[HTML]{FFFFFF}}c 
>{\columncolor[HTML]{FFFFFF}}c 
>{\columncolor[HTML]{FFFFFF}}l 
>{\columncolor[HTML]{FFFFFF}}l 
>{\columncolor[HTML]{FFFFFF}}l 
>{\columncolor[HTML]{FFCCC9}}l }
\hline
\multicolumn{2}{c|}{\cellcolor[HTML]{FFFFFF}} &
  \multicolumn{8}{c}{\cellcolor[HTML]{FFFFFF}w/ Brake ($\mathrm{RMSE}\times10^{-2}\downarrow,\mathrm{R}^2\uparrow$)} \\ \cline{3-10} 
\multicolumn{2}{c|}{\multirow{-2}{*}{\cellcolor[HTML]{FFFFFF}Subject}} &
  \cellcolor[HTML]{FFFFFF}DMD &
  \cellcolor[HTML]{FFFFFF}EDMD &
  \cellcolor[HTML]{FFFFFF}PIDMD &
  \multicolumn{1}{l}{\cellcolor[HTML]{FFFFFF}SUBDMD} &
  \multicolumn{1}{c}{\cellcolor[HTML]{FFFFFF}CSP} &
  \multicolumn{1}{c}{\cellcolor[HTML]{FFFFFF}TFCSP} &
  CTSSP &
  \cellcolor[HTML]{FFFFFF}Proposed \\ \hline
\multicolumn{1}{l|}{\cellcolor[HTML]{FFFFFF}} &
  \texttt{gaa} &
  8.09, 0.74 &
  8.32, 0.72 &
  8.31, 0.72 &
  8.16, 0.73 &
  7.24, 0.79 &
  7.85, 0.75 &
  8.28, 0.72 &
  \textbf{5.80, 0.86} \\
\multicolumn{1}{l|}{\cellcolor[HTML]{FFFFFF}} &
  \texttt{gag} &
  6.78, 0.81 &
  \cellcolor[HTML]{FFFFFF}6.82, 0.81 &
  6.90, 0.81 &
  6.69, 0.82 &
  \textbf{5.96, 0.86} &
  6.48, 0.83 &
  7.00, 0.80 &
  6.45, 0.83 \\
\multicolumn{1}{l|}{\cellcolor[HTML]{FFFFFF}} &
  \texttt{gac} &
  9.33, 0.73 &
  \cellcolor[HTML]{FFFFFF}9.46, 0.73 &
  9.49, 0.72 &
  9.62, 0.72 &
  9.18, 0.74 &
  8.63, 0.77 &
  9.55, 0.72 &
  \textbf{7.47, 0.83} \\
\multicolumn{1}{l|}{\cellcolor[HTML]{FFFFFF}} &
  \texttt{sal} &
  7.16, 0.83 &
  \cellcolor[HTML]{FFFFFF}7.20, 0.83 &
  7.19, 0.83 &
  7.22, 0.83 &
  7.10, 0.84 &
  \textbf{6.92, 0.84} &
  7.19, 0.83 &
  6.95, 0.84 \\
\multicolumn{1}{l|}{\cellcolor[HTML]{FFFFFF}} &
  \texttt{bax} &
  10.0, 0.82 &
  \cellcolor[HTML]{FFFFFF}10.0, 0.82 &
  10.1, 0.82 &
  10.1, 0.82 &
  9.80, 0.83 &
  9.94, 0.82 &
  10.1, 0.82 &
  \textbf{8.37, 0.87} \\
\multicolumn{1}{l|}{\cellcolor[HTML]{FFFFFF}} &
  \texttt{ja} &
  9.68, 0.81 &
  \cellcolor[HTML]{FFFFFF}9.70,0.81 &
  9.81, 0.81 &
  9.79, 0.81 &
  9.61, 0.82 &
  9.67, 0.82 &
  9.80, 0.81 &
  \textbf{8.18, 0.87} \\
\multicolumn{1}{l|}{\cellcolor[HTML]{FFFFFF}} &
  \texttt{bba} &
  4.69, 0.90 &
  \cellcolor[HTML]{FFFFFF}4.64, 0.91 &
  4.72, 0.90 &
  4.74, 0.90 &
  4.69, 0.90 &
  4.88, 0.90 &
  4.76, 0.90 &
  \textbf{4.43, 0.91} \\
\multicolumn{1}{l|}{\cellcolor[HTML]{FFFFFF}} &
  \texttt{ii} &
  5.67, 0.72 &
  \cellcolor[HTML]{FFFFFF}5.68, 0.71 &
  5.74, 0.71 &
  5.70, 0.71 &
  5.64, 0.72 &
  5.60, 0.72 &
  5.83, 0.70 &
  \textbf{4.66, 0.81} \\
\multicolumn{1}{l|}{\cellcolor[HTML]{FFFFFF}} &
  \texttt{gah} &
  7.43, 0.80 &
  \cellcolor[HTML]{FFFFFF}7.50, 0.80 &
  7.53, 0.80 &
  7.48, 0.80 &
  7.53, 0.80 &
  7.48, 0.80 &
  7.61, 0.80 &
  \textbf{6.77, 0.84} \\
\multicolumn{1}{l|}{\cellcolor[HTML]{FFFFFF}} &
  \texttt{gab} &
  4.97, 0.84 &
  \cellcolor[HTML]{FFFFFF}4.96, 0.84 &
  4.99, 0.84 &
  4.94, 0.84 &
  4.98, 0.84 &
  4.76, 0.86 &
  5.03, 0.84 &
  \textbf{4.72, 0.86} \\
\multicolumn{1}{l|}{\cellcolor[HTML]{FFFFFF}} &
  \texttt{ih} &
  10.7, 0.76 &
  \cellcolor[HTML]{FFFFFF}10.8, 0.76 &
  10.9, 0.76 &
  10.8, 0.76 &
  10.7, 0.76 &
  \cellcolor[HTML]{FFFFFF}10.5, 0.77 &
  10.9, 0.75 &
  \textbf{10.5, 0.77} \\
\multicolumn{1}{l|}{\cellcolor[HTML]{FFFFFF}} &
  \texttt{gam} &
  5.90, 0.84 &
  \cellcolor[HTML]{FFFFFF}5.91, 0.84 &
  6.02, 0.83 &
  5.95, 0.84 &
  12.3,-0.39 &
  5.87, 0.84 &
  \multicolumn{1}{c}{\cellcolor[HTML]{FFFFFF}6.12, 0.83} &
  \textbf{5.65, 0.85} \\
\multicolumn{1}{l|}{\cellcolor[HTML]{FFFFFF}} &
  \texttt{gal} &
  6.83, 0.84 &
  \cellcolor[HTML]{FFFFFF}6.82, 0.84 &
  6.91, 0.84 &
  6.89, 0.84 &
  6.84, 0.84 &
  6.81, 0.84 &
  6.92, 0.84 &
  \textbf{6.70, 0.85} \\
\multicolumn{1}{l|}{\cellcolor[HTML]{FFFFFF}} &
  \texttt{saj} &
  8.86, 0.75 &
  \cellcolor[HTML]{FFFFFF}8.83, 0.75 &
  8.89, 0.75 &
  8.85, 0.75 &
  8.77, 0.75 &
  8.51, 0.77 &
  8.94, 0.74 &
  \textbf{8.18, 0.78} \\
\multicolumn{1}{l|}{\cellcolor[HTML]{FFFFFF}} &
  \texttt{ae} &
  9.13, 0.71 &
  \cellcolor[HTML]{FFFFFF}9.18, 0.71 &
  9.20, 0.71 &
  9.20, 0.71 &
  9.18, 0.71 &
  9.07, 0.71 &
  9.52, 0.69 &
  \textbf{8.92, 0.72} \\
\multicolumn{1}{l|}{\cellcolor[HTML]{FFFFFF}} &
  \texttt{dx} &
  6.40, 0.81 &
  \cellcolor[HTML]{FFFFFF}6.58, 0.80 &
  6.57, 0.81 &
  6.56, 0.81 &
  6.46, 0.81 &
  \textbf{6.30, 0.82} &
  6.73, 0.80 &
  6.42, 0.81 \\
\multicolumn{1}{l|}{\multirow{-17}{*}{\cellcolor[HTML]{FFFFFF}O. D.}} &
  \texttt{bad} &
  8.13, 0.78 &
  8.18, 0.78 &
  8.19, 0.78 &
  8.29, 0.77 &
  8.00, 0.79 &
  8.02, 0.79 &
  8.35, 0.77 &
  \textbf{6.89, 0.84} \\ \hline
\multicolumn{1}{l|}{\cellcolor[HTML]{FFFFFF}} &
  1 &
  33.5, 0.72 &
  53.9,-0.12 &
  \multicolumn{1}{l}{\cellcolor[HTML]{FFFFFF}38.5, 0.43} &
  \multicolumn{1}{l}{\cellcolor[HTML]{FFFFFF}36.7, 0.48} &
  48.5, 0.41 &
  44.1, 0.25 &
  44.4, 0.24 &
  {\color[HTML]{111133} \textbf{29.4, 0.79}} \\
\multicolumn{1}{l|}{\cellcolor[HTML]{FFFFFF}} &
  2 &
  41.4,-0.01 &
  24.3, 0.65 &
  \multicolumn{1}{l}{\cellcolor[HTML]{FFFFFF}33.0, 0.36} &
  \multicolumn{1}{l}{\cellcolor[HTML]{FFFFFF}37.8, 0.16} &
  29.5, 0.49 &
  35.7, 0.25 &
  37.1, 0.19 &
  {\color[HTML]{111133} \textbf{24.0, 0.66}} \\
\multicolumn{1}{l|}{\cellcolor[HTML]{FFFFFF}} &
  3 &
  42.7, 0.51 &
  30.7, 0.65 &
  \multicolumn{1}{l}{\cellcolor[HTML]{FFFFFF}42.6, 0.32} &
  \multicolumn{1}{l}{\cellcolor[HTML]{FFFFFF}38.1, 0.46} &
  40.2, 0.56 &
  34.0, 0.57 &
  37.7, 0.46 &
  {\color[HTML]{111133} \textbf{22.9, 0.86}} \\
\multicolumn{1}{l|}{\cellcolor[HTML]{FFFFFF}} &
  4 &
  30.1, 0.43 &
  23.0, 0.67 &
  \multicolumn{1}{l}{\cellcolor[HTML]{FFFFFF}29.4, 0.47} &
  \multicolumn{1}{l}{\cellcolor[HTML]{FFFFFF}29.0, 0.48} &
  26.6, 0.55 &
  29.6, 0.46 &
  31.3, 0.39 &
  {\color[HTML]{111133} \textbf{20.8, 0.73}} \\
\multicolumn{1}{l|}{\cellcolor[HTML]{FFFFFF}} &
  5 &
  23.9, 0.80 &
  22.9, 0.75 &
  \multicolumn{1}{l}{\cellcolor[HTML]{FFFFFF}26.6, 0.67} &
  \multicolumn{1}{l}{\cellcolor[HTML]{FFFFFF}28.9, 0.61} &
  31.8, 0.64 &
  35.3, 0.41 &
  34.6, 0.44 &
  {\color[HTML]{111133} \textbf{21.0, 0.84}} \\
\multicolumn{1}{l|}{\cellcolor[HTML]{FFFFFF}} &
  6 &
  54.9,-1.79 &
  37.3,-0.29 &
  \multicolumn{1}{l}{\cellcolor[HTML]{FFFFFF}35.7,-0.17} &
  \multicolumn{1}{l}{\cellcolor[HTML]{FFFFFF}33.4,-0.03} &
  42.2,-0.65 &
  39.8,-0.46 &
  40.9,-0.55 &
  {\color[HTML]{111133} \textbf{32.2, 0.04}} \\
\multicolumn{1}{l|}{\cellcolor[HTML]{FFFFFF}} &
  7 &
  28.1, 0.55 &
  27.9, 0.48 &
  \multicolumn{1}{l}{\cellcolor[HTML]{FFFFFF}26.8, 0.52} &
  \multicolumn{1}{l}{\cellcolor[HTML]{FFFFFF}25.8, 0.56} &
  35.5, 0.29 &
  31.4, 0.35 &
  28.6, 0.46 &
  {\color[HTML]{111133} \textbf{19.8, 0.78}} \\
\multicolumn{1}{l|}{\cellcolor[HTML]{FFFFFF}} &
  8 &
  25.5, 0.69 &
  29.2, 0.51 &
  \multicolumn{1}{l}{\cellcolor[HTML]{FFFFFF}27.9, 0.55} &
  \multicolumn{1}{l}{\cellcolor[HTML]{FFFFFF}27.7, 0.56} &
  35.9, 0.39 &
  32.7, 0.39 &
  30.9, 0.46 &
  {\color[HTML]{111133} \textbf{21.9, 0.77}} \\
\multicolumn{1}{l|}{\cellcolor[HTML]{FFFFFF}} &
  9 &
  33.7, 0.31 &
  30.4, 0.43 &
  \multicolumn{1}{l}{\cellcolor[HTML]{FFFFFF}36.9, 0.16} &
  \multicolumn{1}{l}{\cellcolor[HTML]{FFFFFF}35.1, 0.24} &
  42.2,-0.09 &
  34.5, 0.27 &
  34.4, 0.27 &
  {\color[HTML]{111133} \textbf{24.5, 0.63}} \\
\multicolumn{1}{l|}{\cellcolor[HTML]{FFFFFF}} &
  10 &
  34.8, 0.49 &
  29.4, 0.59 &
  \multicolumn{1}{l}{\cellcolor[HTML]{FFFFFF}34.9, 0.43} &
  \multicolumn{1}{l}{\cellcolor[HTML]{FFFFFF}39.7, 0.27} &
  40.1, 0.32 &
  41.9, 0.18 &
  44.7, 0.06 &
  {\color[HTML]{111133} \textbf{29.4, 0.63}} \\
\multicolumn{1}{l|}{\cellcolor[HTML]{FFFFFF}} &
  11 &
  34.6, 0.24 &
  24.9, 0.73 &
  \multicolumn{1}{l}{\cellcolor[HTML]{FFFFFF}29.9, 0.31} &
  \multicolumn{1}{l}{\cellcolor[HTML]{FFFFFF}29.7, 0.32} &
  36.7, 0.14 &
  32.4, 0.19 &
  34.1, 0.11 &
  {\color[HTML]{111133} \textbf{20.2, 0.74}} \\
\multicolumn{1}{l|}{\cellcolor[HTML]{FFFFFF}} &
  12 &
  55.8,-0.14 &
  55.1,-0.11 &
  \multicolumn{1}{l}{\cellcolor[HTML]{FFFFFF}58.8,-0.27} &
  \multicolumn{1}{l}{\cellcolor[HTML]{FFFFFF}56.2,-0.16} &
  53.2,-0.04 &
  46.0, 0.22 &
  47.2, 0.18 &
  {\color[HTML]{111133} \textbf{21.2, 0.62}} \\
\multicolumn{1}{l|}{\cellcolor[HTML]{FFFFFF}} &
  13 &
  37.9, 0.59 &
  37.3, 0.50 &
  \multicolumn{1}{l}{\cellcolor[HTML]{FFFFFF}40.3, 0.42} &
  33.8, 0.59 &
  \textbf{27.7, 0.78} &
  47.9, 0.18 &
  45.6, 0.25 &
  {\color[HTML]{111133} 35.0, 0.65} \\
\multicolumn{1}{l|}{\multirow{-14}{*}{\cellcolor[HTML]{FFFFFF}H. S.}} &
  14 &
  37.6, 0.57 &
  33.4, 0.60 &
  \multicolumn{1}{l}{\cellcolor[HTML]{FFFFFF}38.4, 0.47} &
  33.6, 0.60 &
  45.4, 0.37 &
  33.3, 0.60 &
  45.4, 0.27 &
  {\color[HTML]{111133} \textbf{27.6, 0.77}} \\ \hline
\end{tabular}
\end{table*}

\section{Experiments}
\subsection{Comparative Experiments}

We evaluated seven baselines for extracting features from the preprocessed $59$-electrode EEG signals.

\subsubsection{Common Spatial Patterns (CSP)}
CSP algorithm is a well-established technique for extracting discriminative features in binary-class EEG classification tasks, such as motor imagery decoding \cite{jiang2020efficient}. CSP identifies spatial filters that maximize the variance ratio between two classes (e.g., braking and non-braking). Despite its effectiveness, CSP operates on a fixed time window and assumes signals' stationarity, limiting its ability to capture dynamic neurophysiological activity. Moreover, its reliance on a predefined frequency band may discard discriminative spectral information. To address these issues, Time-Frequency Common Spatial Patterns (TFCSP) algorithm \cite{mishuhina2021complex} decomposes EEG signals into multiple overlapping sub-bands and applies CSP to each, thereby capturing spectral characteristics while reducing sensitivity to frequency misalignment. Further, the Common Temporal-Spectral-Spatial Patterns (CTSSP) algorithm \cite{pan2025ctssp} introduces multi-time-window modeling and joint temporal-spectral-spatial optimization to adaptively extract features reflecting distinct neurophysiological activity, enhancing robustness to EEG non-stationarity.

We applied CSP and its variants to the EEG segments during braking versus non-braking conditions. The spatial filters corresponding to the top three eigenvalues were retained, consistent with standard CSP implementations \cite{Mousavi2022SpectrallyAC}. The prediction model integrates these CSP inputs as follows:

\begin{equation}
y_{t+\Delta t}^{\mathrm{pre}}=\mathrm{MLP}      (\mathbf{p}^{\mathrm{csp}},\mathbf{y}^{\mathrm{his}})
\label{csp}
\end{equation}
where $\mathbf{p}^{\mathrm{csp}}=[p_{t-(m-1)h:t}^{\mathrm{csp_1}},\dots,p_{t-(m-1)h:t}^{\mathrm{csp_3}}]$. To ensure comparability, we computed the $\mathbf{p}^{\mathrm{csp}}$ of TFCSP and CTSSP as detailed in original works \cite{mishuhina2021complex,pan2025ctssp}. Moreover, we calculated the power within identified band as the $\mathbf{p}^{\mathrm{csp}}$ of CSP.

\subsubsection{Dynamic Mode Decomposition (DMD)}

Since EEG signals originate from high-dimensional dynamical systems \cite{cura2021analysis}, we employed DMD \cite{kutz2016dynamic} to extract spatiotemporal dynamics by approximating the system as linear. DMD decomposes EEG signals into spatial modes and their associated temporal dynamics, thereby enabling a compact representation of oscillatory behavior. However, DMD suffers from three key limitations: (i) Its linearity assumption contradicts the nonlinear nature of EEG dynamics; (ii) its data-driven nature may violate physical laws; and (iii) its eigenvalue estimates are noise-sensitive. To mitigate these issues, we adopt three DMD extensions: (i) Extended Dynamic Mode Decomposition (EDMD) \cite{williams2015data}, which embeds state variables into a nonlinear feature space for linear approximation; (ii) Physics-informed DMD (PIDMD) \cite{baddoo2023physics}, which enforces physical constraints (e.g., symmetry or causality) during optimization to improve interpretablility; and (iii) Subspace DMD (SUBDMD) \cite{takeishi2017subspace}, which reduces noise by projecting EEG data onto a lower-dimensional subspace. We benchmarked DMD and its three extensions (EDMD, PIDMD, and SUBDMD). To enhance temporal modeling, EEG signals were embedded into Hankel matrices before decomposition. The DMD spectrum \cite{dmdspectrum}, computed from the Koopman operator's eigendecomposition, served as the feature for prediction:

\begin{equation}
y_{t+\Delta t}^{\mathrm{pre}}=\mathrm{MLP}      (\mathbf{p}^{\mathrm{dmd}},\mathbf{y}^{\mathrm{his}})
\label{dmd}
\end{equation}
where $\mathbf{p}^{\mathrm{dmd}}=p_{t-(m-1)h:t}^{\mathrm{dmd}}$ encodes the time-lagged DMD power spectrum.

\subsection{Ablation Experiments}
To evaluate the contributions of EEG features and historical braking intensity ($\mathbf{y}^{\mathrm{his}}$), we conducted three ablation studies: 
\begin{itemize}
    \item Only $\mathbf{y}^{\mathrm{his}}$ was used to assess its standalone predictive performance.
    \item $\mathbf{y}^{\mathrm{his}}$ was excluded to quantify the impact of EEG features extracted by our proposed method and baselines.
    \item Braking-related ICs were replaced by EEG signals to evaluate the effectiveness of our IC selection strategy.
\end{itemize}

Prediction performance is quantified using the $\mathrm{RMSE}$ and the coefficient of determination (i.e., R-Square, $\mathrm{R}^2$):

\begin{equation}
\mathrm{RMSE}=\sqrt{\frac{1}{N}\sum_{t=1}^N{(y_{t+\Delta t}^{\mathrm{tru}}-y_{t+\Delta t}^{\mathrm{pre}})}^2},\label{rmse}
\end{equation}

\begin{equation}
\mathrm{R}^2=1-\frac{\sum_{t=1}^N{(y_{t+\Delta t}^{\mathrm{tru}}-y_{t+\Delta t}^{\mathrm{pre}})}}{\sum_{t=1}^N{(y_{t+\Delta t}^{\mathrm{tru}}-\bar{y}_{t+\Delta t}^{\mathrm{tru}})}}
\label{R}
\end{equation}
where $N$ is the sample size, $y_{t+\Delta t}^{\mathrm{tru}}$, $y_{t+\Delta t}^{\mathrm{pre}}$ are the ground truth and predicted braking intensity at time $t+\Delta t$, and $\bar{y}_{t+\Delta t}^{\mathrm{tru}}$ is the mean of the true values. Smaller RMSE and higher $\mathrm{R}^2$ indicate better performance.

\definecolor{lightgreen}{rgb}{0.5, 0.99, 0.5}
\definecolor{lightblue}{RGB}{200, 200, 255}

\subsection{Results}
\begin{figure}[t]
    \centering
    \includegraphics[width=0.47\textwidth]{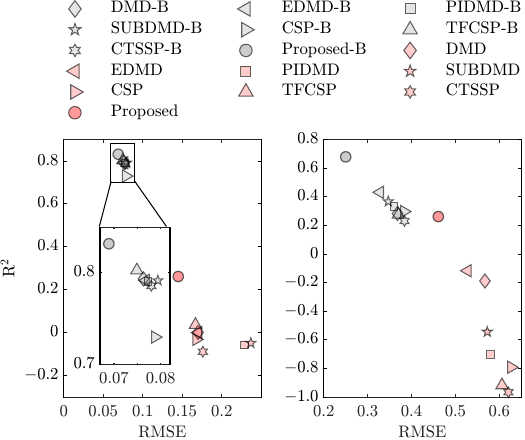}
    \caption{The dataset-level prediction performance with (red) and without (gray) braking intensity as input in (left) O.D. and (right) H.S..}
    \label{multi_methods}
\end{figure} 
We evaluate the proposed method with comparative experiments and ablation studies on two datasets. Table \ref{table1} compares the subject-level performance of our method with baselines with braking intensity as input, where the best and our method's results are highlighted in $\mathbf{bold}$ and red, respectively. 

Our method achieves the highest accuracy on $14/17$ subjects in O.D. and $13/14$ subjects in H.S., with $\mathrm{RMSE}$ reductions of up to $19.9\%$ and $53.9\%$, respectively, demonstrating superior generalizability. Baseline methods show inconsistent performance. For instance, TFCSP performs best among baselines in O.D. (outperforming ours on two subjects) but only surpasses other baselines on $3$ subjects in H.S. EDMD leads baselines in H.S. but underperforms in O.D.

Fig. \ref{multi_methods} compares the dataset-level prediction accuracy of our method with (red) and without (gray) braking intensity as input.

\textbf{The proposed method outperforms baselines}. Across both datasets, our method achieves the highest performance, with significantly improved $\mathrm{RMSE}$ and $\mathrm{R}^2$ values, regardless of whether braking intensity is included. When braking intensity is included, our method surpasses the best baselines by substantial margin: $\mathrm{R}^2$ increases by $3.6\%$ (O.D.) and $57.5\%$ (H.S.), while $\mathrm{RMSE}$ decreases by $8.0\%$ (O.D.) and $23.8\%$ (H.S.). Notably, even without braking intensity as input, our method maintains significant advantages. Compared to the best baselines, $\mathrm{RMSE}$ decreases by $13.1\%$ (O.D.) and $12.6\%$ (H.S.), and $\mathrm{R}^2$ increases from $0.00$ (O.D.) and $-0.12$ (H.S.) to $0.26$ (O.D.) and $0.08$ (H.S.). These results highlight the robustness of our approach regardless of input configuration.

\begin{figure}
\centering
\centerline{\includegraphics[width=0.47\textwidth]{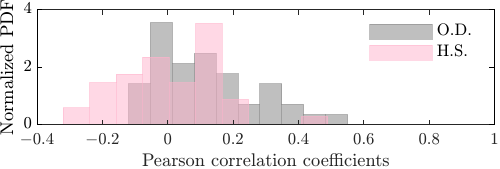}}
\caption{Histograms of Pearson correlation coefficients between power peaks and braking intensity peaks across all subjects in O.D. and H.S..}
\label{SCATTER}
\end{figure}

\newcolumntype{C}{>{\centering\arraybackslash}X}  % 居中X列
\newcolumntype{Y}{>{\centering\arraybackslash}p{0.8cm}}
\begin{table}[]
\caption{Results of Data Ablation Experiments for Two Datasets with the Best Result Marked in $\mathbf{bold}$. The $\mathrm{RMSE}$ Values are Reported in Units of $10^{-2}$.}
\label{dataablation}
\centering
\begin{tabularx}{\linewidth}{c|YYY|CC|CC}
\hline
                           & \multicolumn{3}{c|}{Data} & \multicolumn{2}{c|}{O.D.}     & \multicolumn{2}{c}{H.S.}      \\ \cline{2-8} 
\multirow{-2}{*}{Method}   & Brake    & EEG     & ICs   & $\mathrm{RMSE}$          & $\mathrm{R}^{2}$             & $\mathrm{RMSE}$           & $\mathrm{R}^{2}$            \\ \hline
                           & \CheckmarkBold       &        &       & 7.69          & 0.79          & 27.7          & 0.69          \\
\multirow{-2}{*}{Contrast} & \CheckmarkBold       & \CheckmarkBold     &       & 7.24          & 0.82          & 31.4          & 0.56          \\
\rowcolor[HTML]{FFCCC9} 
Ours                       & \CheckmarkBold       &        & \CheckmarkBold    & \textbf{6.89} & \textbf{0.83} & \textbf{25.0} & \textbf{0.74} \\ \hline
\end{tabularx}
\end{table}

\textbf{Braking intensity enhances prediction accuracy}. Omitting braking intensity yields significant performance degradation in both datasets. For instance, our method exhibits a $110.3\%$ increase in $\mathrm{RMSE}$ (O.D.) and a $84.4\%$ increase (H.S.), with $\mathrm{R}^2$ decreasing by $68.5\%$ (O.D.) and $87.7\%$ (H.S.). Compared with our method, baselines show even larger declines. For instance, EDMD exhibits a $121.6\%$ rise in $\mathrm{RMSE}$ (O.D.) and $60.9\%$ (H.S.), with $\mathrm{R}^2$ dropping below zero in both datasets. This degradation is further analyzed in Fig. \ref{SCATTER}, which examines the correlation between power peak values and braking intensity peak values using Pearson coefficients. Most subjects show weak correlations ($-0.4$ to $0.6$), with none exceeding $0.8$ in either datasets. These results align with prior findings \cite{vecchiato2019electroencephalographic}, confirming that power peaks alone cannot reliably predict braking intensity magnitude. While power data capture temporal trends, accurate prediction requires combining both features: power for trend timing and braking intensity for magnitude scaling.

Table \ref{dataablation} presents the data ablation results on two datasets, with the best performance in $\mathbf{bold}$ and our method's results in red. The `EEG' corresponds to the prediction using $59$-electrode EEG signals. We computed the correlation metric $\nu$ for each electrode and selected the top $5\%$ electrodes based on $\nu$ values as inputs. Our proposed method significantly outperforms its counterparts by integrating braking-related ICs with braking data. These results demonstrate that our source selection strategy effectively captures neurophysiological activity related to braking.

\begin{figure}[t]
    \centering
    \includegraphics[width=0.48\textwidth]{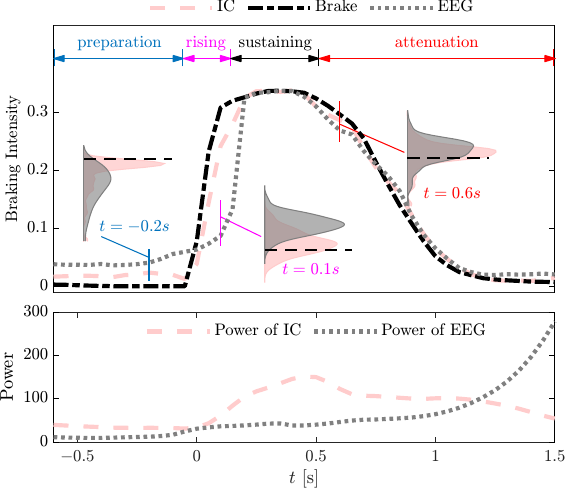}
    \caption{Top: Actual and predicted braking intensity for subject \texttt{gaa} using the braking-related IC versus preprocessed EEG inputs, with error distributions of three moments. Bottom: Power trajectories of the braking-related IC and EEG.}
    \label{fig12}
\end{figure}

To explain the superior performance of our method, we compares the braking intensity prediction based on the braking-related IC (IC$37$) and EEG data for subject \texttt{gaa} of O.D., along with their power trajectories (Fig. \ref{fig12}). Predictions from EEG data exhibit significant deviations from the ground truth. To analyze these deviations, we divide the braking process into four phases: \textcolor[rgb]{0,0.4470,0.7410}{preparation}, \textcolor[rgb]{1,0,1}{rising}, sustaining, and \textcolor{red}{attenuation}. During the \textcolor[rgb]{0,0.4470,0.7410}{preparation} phase (no actual deceleration), the EEG data incorrectly predicts non-zero values. Additionally, its prediction lags behind the ground truth in the \textcolor[rgb]{1,0,1}{rising} phase and shows slight temporal misalignment in the \textcolor{red}{attenuation} phase. In contrast, predictions from the braking-related IC closely match the ground truth, overcoming these limitations. 

The error distributions (residuals between predicted and actual values) for the three braking phases are shown in Fig. \ref{fig12}. Pink and gray distributions represent errors from the braking-related IC and EEG data, respectively. Errors from the braking-related IC are tightly clustered near zero across all phases, demonstrating high accuracy, whereas EEG data yields errors with higher variance and bias. This performance gap arises because the power trajectory of braking-related IC exhibits temporal dynamics and high similarity to the ground-truth braking intensity. In contrast, the power trajectory of EEG displays greater variability and weaker behavioral correlation, likely due to contamination by artifacts and non-braking-related neural activity.

\section{Conclusion}

This paper presented an electroencephalography (EEG)-based framework to predict emergency braking intensity. The method isolates braking-related independent components (ICs) through decoupling neural signatures from artifact interference, demonstrating superior predictive performance over state-of-the-art algorithms in comparative experiments. Hierarchical clustering identifies two distinct scalp topographies among braking-related ICs, with (\texttt{IC-FPO} and \texttt{IC-C}) showing cross-subject consistency. \texttt{IC-FPO} exhibits a frontal-parietal-occipital dipole distribution (negative frontal and positive parietal-occipital potentials), while \texttt{IC-C} displays prominent central negativity. Such topographical consistency implies common neural mechanisms for braking intention generation. In human-in-the-loop simulation, these ICs show trial-invariant temporal patterns, suggesting stable neural substrates underlying emergency braking behavior. Our findings highlight the feasibility of using EEG-derived ICs to decode braking intentions in human-machine systems, advancing neural interfaces in vehicular control.

\appendices
\section*{Appendix A}
The power spectral density (PSD) quantifies the distribution of signal power across frequency and underlies band power estimation. To compute the PSD, we employed Welch’s method \cite{welch1967use}, which involves four key steps: segmentation, windowing, periodogram computation, and averaging. The EEG signals were divided into overlapping segments. Then, a Hamming window was applied to each segment to reduce spectral leakage. The windowed segments were subsequently transformed into the frequency domain via the discrete Fourier transform, where the squared magnitude of the Fourier coefficients gave the periodogram. Finally, these periodograms were averaged across segments to produce the final PSD estimation. Band power was calculated by integrating the PSD over the target frequency range. For discrete spectra, the integral was approximated by summing the PSD values within the band multiplied by the frequency bin width.

\section*{Appendix B}
The hyperparameters referenced in Section \uppercase\expandafter{\romannumeral3}-B were selected as follows.

First, we analyzed the distribution of braking durations across all trials to determine the time interval. A $1.5$-s interval covers $96.9\%$ of trials, indicating its adequacy for capturing dynamic characteristics. Thus, we selected a symmetric $3$-s window ($\pm1.5$ s) centered at braking onset, with boundaries at $t=-1.5$ s (left) and $t=1.5$ s (right).

Second, the window length $w$ for $\delta$-$\theta$ band power ($0.5$-$8$ Hz) was chosen to balance temporal and frequency resolution. For a signal sampled at frequency $f_s$ with $N$ points, the window duration is $T = N/f_s$. From Fourier transform principles, the frequency resolution $\Delta f$ is:

\begin{equation}
    \Delta f = \frac{f_s}{N}=\frac{1}{T}
\label{deltaf}
\end{equation}
For instance, a $1000$-ms window yields $\Delta f=1$ Hz (high frequency resolution), and a $250$-ms window yields $\Delta f=4$ Hz (poorer frequency resolution but higher temporal resolution). To reconcile this trade-off, we select $T = 500$ ms, providing adequate resolution in both domains.

Third, to resolve rapid transients in $0.5$-$8$ Hz EEG signals, the step size must be shorter than half the period of the highest frequency, i.e., $8$ Hz ($62.5$ ms). Thus, we chose a step size of $50$ ms to ensure precise temporal tracking while maintaining computational efficiency.

% \bibliographystyle{IEEEtran}
% \bibliographystyle{ieeetr}
% \bibliography{sample}

\begin{thebibliography}{1}
\bibliographystyle{IEEEtran}

\bibitem{wang2022effect}
X. Wang, X. Zhang, F. Guo, Y.Gu, and X. Zhu, ``Effect of daily car-following behaviors on urban roadway rear-end crashes and near-crashes$\colon$A naturalistic driving study,'' \textit{Accident Anal. Prevention}, vol. 164, Jan. 2022, Art. no. 106502.

% \bibitem{zheng2014study}
% R. Zheng, K. Nakano, S. Yamabe, M. Aki, H. Nakamura, and Y. Suda, ``Study on emergency-avoidance braking for the automatic platooning of trucks,'' \textit{IEEE Trans. Intell. Transp. Syst.}, vol. 15, no. 4, pp. 1748-–1757, Aug. 2014.

\bibitem{lyu2020towards}
F. Lyu, N. Cheng, H. Zhu, H. Zhou, W. Xu, M. Li, and X. Shen, ``Towards rear-end collision avoidance$\colon$Adaptive beaconing for connected vehicles,'' \textit{IEEE Trans. Intell. Transp. Syst.}, vol. 22, no. 2, pp. 1248-–1263, Feb. 2021.

% \bibitem{lee2016real}
% D. Lee and H. Yeo, ``Real-time rear-end collision-warning system using a multilayer perceptron neural network,'' \textit{IEEE Trans. Intell. Transp. Syst.}, vol. 17, no. 11, pp. 3087-–3097, Nov. 2016.

\bibitem{gunjate2023systematic}
S. S. Gunjate and S. A. Khot, ``A systematic review of emergency braking assistant system to avoid accidents using pulse width modulation and fuzzy logic control integrated with antilock braking,'' \textit{Int. J. Automot. Mech. Eng.}, vol. 20, no. 2, pp. 10457-–10479, Jul. 2023.

\bibitem{kusano2012safety}
K. D. Kusano and H. C. Gabler, ``Safety benefits of forward collision warning, brake assist, and autonomous braking systems in rear-end collisions,'' \textit{IEEE Trans. Intell. Transp. Syst.}, vol. 13, no. 4, pp. 1546-–1555, Dec. 2012.

\bibitem{muller2008machine}
K.-R. M\"{u}ller, M. Tangermann, G. Dornhege, M. Krauledat, G. Curio, and B. Blankertz, ``Machine learning for real-time single-trial eeg analysis$\colon$from brain–computer interfacing to mental state monitoring,'' \textit{J. Neurosci. Methods}, vol. 167, no. 1, pp. 82-–90, Jan. 2008.

\bibitem{teng2017eeg}
T. Teng, L. Bi, and Y. Liu, ``Eeg-based detection of driver emergency braking intention for brain-controlled vehicles,'' \textit{IEEE Trans. Intell. Transp. Syst.}, vol. 19, no. 6, pp. 1766-–1773, Jun. 2018.

\bibitem{ju2022recognition}
J. Ju, A. G. Feleke, L. Luo, and X. Fan, ``Recognition of drivers’ hard and soft braking intentions based on hybrid brain-computer interfaces,'' \textit{Cyborg Bionic Syst.}, Jan. 2022.

\bibitem{wang2017eeg}
H. Wang, L. Bi, and T. Teng, ``Eeg-based emergency braking intention prediction for brain-controlled driving considering one electrode falling off,'' in \textit{Proc. Annu. Int. Conf. IEEE Eng. Med. Biol. Soc. (EMBS)}, Jul. 2017, pp. 2494-–2497.

\bibitem{kim2014detection}
I.-H. Kim, J.-W. Kim, S. Haufe, and S.-W. Lee, ``Detection of braking intention in diverse situations during simulated driving based on eeg feature combination,'' \textit{J. Neural Eng.}, vol. 12, no. 1, p. 016001, Nov. 2014.

\bibitem{liang2023eeg}
X. Liang, Y. Yu, Y. Liu, K. Liu, Y. Liu, and Z. Zhou, ``Eeg-based emergency braking intention detection during simulated driving,'' \textit{Biomed. Eng. Online}, vol. 22, no. 1, p. 65, Jul. 2023.

\bibitem{mora2023simplified}
H. J. Mora and E. J. Pino, ``Simplified prediction method for detecting the emergency braking intention using eeg and a cnn trained with a 2d matrices tensor arrangement,'' \textit{Int. J. Hum.-Comput. Interact.}, vol. 39, no. 3, pp. 587-–600, Apr. 2023.

\bibitem{lutes2024convolutional}
N. Lutes, V. S. S. Nadendla, and K. Krishnamurthy, ``Convolutional spiking neural networks for intent detection based on anticipatory brain potentials using electroencephalogram,'' \textit{Sci. Rep.}, vol. 14, no. 1, p. 8850, Apr. 2024.

\bibitem{wenzhuo2025cvpr}
W. Liu et al., ``MMTL-UniAD$\colon$ A unified framework for multimodal and multi-task learning in assistive driving perception,'' 2025, \textit{arXiv $\colon$2504.02264}.

\bibitem{wenzhuo2026}
W. Liu et al., ``UV-M3TL$\colon$ A unified and versatile multimodal multi-task learning framework for assistive driving perception,'' 2026, \textit{arXiv $\colon$2602.01594}.

\bibitem{kim2014decision}
J.-W. Kim, I.-H. Kim, and S.-W. Lee, ``Decision of braking intensity during simulated driving based on analysis of neural correlates,'' in \textit{Proc. IEEE Int. Conf. Syst. Man Cybern. (SMC)}, Apr. 2014, pp. 4129–-4132.

\bibitem{zhang2025generative}
F. Zhang, L. Dong, L. Zhang, X. Zhang, and X. Jiao, ``Generative AI-Enhanced Multi-Physiological Signal Analysis for Intelligent Transportation Safety$\colon$An Attention-Masked Transformer Approach,'' \textit{IEEE Trans. Intell. Transp. Syst.}, early access, pp. 1--12, 2025, doi$\colon$10.1109/TITS.2025.3554808.

\bibitem{li2025assessing}
P. Li, G. Qi, S. Zhao, and W. Guan, ``Assessing the effects of artifacts and noise in eeg signals on car-following driving behavior prediction,'' \textit{Biomed. Signal Process. Control}, vol. 100, p. 106922, Feb. 2025.

\bibitem{bss}
M. Wang, C. Ma, Z. Li, S. Zhang, and Y. Li, ``Alertness Estimation Using Connection Parameters of the Brain Network,'' \textit{IEEE Trans. Intell. Transp. Syst.}, vol. 23, no. 12, pp. 25448--25457, Dec. 2022.

\bibitem{luck2014introduction}
S. J. Luck, \textit{An introduction to the event-related potential technique}. MIT press, 2014.

\bibitem{stevens2019creativity}
C. E. Stevens Jr and D. L. Zabelina, ``Creativity comes in waves: an eeg-focused exploration of the creative brain,'' \textit{Curr. Opin. Behav. Sci.}, vol. 27, pp. 154--162, Jun. 2019.

\bibitem{ouyang2020decomposing}
G. Ouyang, A. Hildebrandt, F. Schmitz, and C. S. Herrmann, ``Decomposing alpha and 1/f brain activities reveals their differential associations with cognitive processing speed,'' \textit{NeuroImage}, vol. 205, p. 116304, Jan. 2020.

\bibitem{morales2022time}
S. Morales and M. E. Bowers, ``Time-frequency analysis methods and their application in developmental eeg data,'' \textit{Dev. Cognit. Neurosci.}, vol. 54, p. 101067, Apr. 2022.

\bibitem{vecchio2022time}
F. Vecchio, L. Nucci, C. Pappalettera, F. Miraglia, D. Iacoviello, and P. M. Rossini, ``Time-frequency analysis of brain activity in response to directional and non-directional visual stimuli$\colon$an event related spectral perturbations (ersp) study,'' \textit{J. Neural Eng.}, vol. 19, no.6, p. 066004, Nov. 2022.

\bibitem{makeig2011erp}
S. Makeig, J. Onton, et al., ``Erp features and eeg dynamics$\colon$an ica perspective,'' \textit{Oxford handbook of event-related potential components}, pp. 51–86, 2011.

\bibitem{kraskov2004estimating}
A. Kraskov, H. St\"{o}gbauer, and P. Grassberger, ``Estimating mutual information,'' \textit{Phys. Rev. E: Stat. Nonlinear Soft Matter Phys.}, vol. 69, no.6, p. 066138, Jun. 2004.

\bibitem{bell1995information}
A. J. Bell and T. J. Sejnowski, ``An information-maximization approach to blind separation and blind deconvolution,'' \textit{Neural Comput.}, vol. 7, no.6, pp. 1129--1159, Nov. 1995.

\bibitem{papoulis1965random}
A. Papoulis, \textit{Random variables and stochastic processes}. New York, NY, USA$\colon$McGraw Hill, 1965.

\bibitem{bhattacharyya2022ocular}
A. Bhattacharyya, A. Verma, R. Ranta, and R. B. Pachori, ``Ocular artifacts elimination from multivariate eeg signal using frequency-spatial filtering,'' \textit{IEEE Trans. Cognit. Dev. Syst.}, vol. 15, no.3, pp. 1547--1559, Sep. 2022.

\bibitem{pion2019iclabel}
L. Pion-Tonachini, K. Kreutz-Delgado, and S. Makeig, ``Iclabel: An automated electroencephalographic independent component classifier, dataset, and website,'' \textit{NeuroImage}, vol. 198, pp. 181--197, Sep. 2019.

\bibitem{pancholi2022source}
S. Pancholi, A. Giri, A. Jain, L. Kumar, and S. Roy, ``Source aware deep learning framework for hand kinematic reconstruction using eeg signal,'' \textit{IEEE Trans. Cybern.}, vol. 53, no. 7, pp. 4094--4106, Jul. 2023.

\bibitem{he2015delving}
K. He, X. Zhang, S. Ren, and J. Sun, ``Delving deep into rectifiers$\colon$Surpassing human-level performance on imagenet classification,'' in \textit{Proc. IEEE Int. Conf. Comput. Vis. (ICCV)}, Dec. 2015, pp. 1026–-1034.

\bibitem{kingma2014adam}
D. P. Kingma, ``Adam$\colon$A method for stochastic optimization,'' 2014, \textit{arXiv $\colon$1412.6980}.

\bibitem{haufe2011eeg}
S. Haufe, M. S. Treder, M. F. Gugler, M. Sagebaum, G. Curio, and B. Blankertz, ``Eeg potentials predict upcoming emergency brakings during simulated driving,'' \textit{J. Neural Eng.}, vol. 8, no. 5, p. 056001, Jul. 2011.

\bibitem{gao2025cross}
D. Gao, X. Tao, X. Wu, B. Du, Y. Qin, and J. Lu, ``Cross-scenario vigilance detection based on eeg analysis for safety driving in autonomous,'' \textit{IEEE Trans. Intell. Transp. Syst.}, vol. 26, no. 8, pp. 11404--11419, Aug. 2025.

\bibitem{hu2024eeg}
F. Hu, L. Zhang, X. Yang, and W. -A. Zhang, ``EEG-based driver fatigue detection using spatio-temporal fusion network with brain region partitioning strategy,'' \textit{IEEE Trans. Intell. Transp. Syst.}, vol. 25, no. 8, pp. 9618--9630, Aug. 2024.

\bibitem{palaniappan2021investigating}
R. Palaniappan, S. Mouli, H. Bowman, and I. McLoughlin, ``Investigating the cognitive response of brake lights in initiating braking action using eeg,'' \textit{IEEE Trans. Intell. Transp. Syst.}, vol. 23, no. 8, pp. 13878--13883, Aug. 2022.

\bibitem{lyu2025driver}
X. Lyu, M. Azeem Akbar, S. Manimurugan, and H. Jiang, ``Driver Fatigue Warning Based on Medical Physiological Signal Monitoring for Transportation Cyber-Physical Systems,'' \textit{IEEE Trans. Intell. Transp. Syst.}, vol. 26, no. 9, pp. 14237--14249, Sep. 2025.

\bibitem{delorme2004eeglab}
A. Delorme and S. Makeig, ``Eeglab: an open source toolbox for analysis of single-trial eeg dynamics including independent component analysis,'' \textit{J. Neurosci. Methods}, vol. 134, no. 1, pp. 9--21, Mar. 2004.

\bibitem{gewers2021principal}
F. L. Gewers, G. R. Ferreira, H. F. D. Arruda, F. N. Silva, C. H. Comin, D. R. Amancio, and L. d. F. Costa, ``Principal component analysis$\colon$A natural approach to data exploration,'' \textit{ACM Comput. Surv.}, vol. 54, no. 4, pp. 1--34, May 2021.

\bibitem{sokal1958statistical}
R. R. Sokal, \textit{et al.}, ``A statistical method for evaluating systematic relationships,'' \textit{J. Neurosci. Methods}, vol. 134, no. 1, pp. 9--21, Mar. 2004.

\bibitem{cos2022}
A. Li, C. Fan, F. Xiao, and Z. Chen, ``Distance measures in building informatics$\colon$An in-depth assessment through typical tasks in building energy management,'' \textit{Energy Build.}, vol. 258, Mar. 2022, Art. no.111817.

\bibitem{leon1998end}
O. Leon Nehmad, ``The end in sight$\colon$A look at the occipital lobe,'' \textit{Clin. Eye Vision Care}, vol. 10, no. 3, pp. 125--133, Sep. 1998.

\bibitem{thier1997parietal}
P. Thier, H.-O. Karnath, \textit{Parietal lobe contributions to orientation in 3D space}. Heidelberg, Germany$\colon$Springer Verlag, 1997.

\bibitem{schall2002monitoring}
J. D. Schall, V. Stuphorn, and J. W. Brown, ``Monitoring and control of action by the frontal lobes,'' \textit{Neuron}, vol. 36, no. 2, pp. 309--322, Oct. 2002.

\bibitem{praamstra2005frontoparietal}
P. Praamstra, L. Boutsen, and G. W. Humphreys, ``Frontoparietal control of spatial attention and motor intention in human eeg,'' \textit{J. Neurophysiol.}, vol. 94, no. 1, pp. 764--774, Jul. 2005.

\bibitem{naito2004sensing}
E. Naito, ``Sensing limb movements in the motor cortex: how humans sense limb movement,'' \textit{The Neuroscientist}, vol. 10, no. 1, pp. 73--82, Feb. 2004.

\bibitem{howmanyelectrodes}
T. M. Lau, J. T. Gwin, and D. P. Ferris, ``How many electrodes are really needed for EEG-based mobile brain imaging?'' \textit{J. Behav. Brain Sci.}, vol. 2, 2012, Art. no. 22107.

\bibitem{jiang2020efficient}
A. Jiang, J. Shang, X. Liu, Y. Tang, H. K. Kwan, and Y. Zhu, ``Efficient csp algorithm with spatio-temporal filtering for motor imagery classification,'' \textit{IEEE Trans. Neural Syst. Rehabil. Eng.}, vol. 28, pp. 1006--1016, Apr. 2020.

\bibitem{mishuhina2021complex}
V. Mishuhina, and X. Jiang, ``Complex common spatial patterns on time-frequency decomposed EEG for brain-computer interface,'' \textit{Pattern Recognit.}, vol. 115, p. 107918, Jul. 2021.

\bibitem{pan2025ctssp}
L. Pan, K. Wang, W. Yi, Y. Zhang, M. Xu, and D. Ming, ``CTSSP$\colon$ A Temporal-Spectral-Spatio Joint Optimization Algorithm for Motor Imagery EEG Decoding,'' \textit{Authorea Preprints}, Apr. 2025.

\bibitem{Mousavi2022SpectrallyAC}
M. Mousavi, E. Lybrand, S. Feng, S. Tang, R. Saab, and Virginia R. de Sa, ``Spectrally Adaptive Common Spatial Patterns,'' 2022, \textit{arXiv $\colon$ 2202.04542}.

\bibitem{cura2021analysis}
O. K. Cura and A. Akan, ``Analysis of epileptic eeg signals by using dynamic mode decomposition and spectrum,'' \textit{Biocybern. Biomed. Eng.}, vol. 41, no. 1, pp. 28--44, Jan. 2021.

\bibitem{kutz2016dynamic}
J. N. Kutz, S. L. Brunton, B. W. Brunton, and J. L. Proctor, \textit{Dynamic mode decomposition: data-driven modeling of complex systems}. Philadelphia, PA, USA$\colon$ SIAM, 2016.

\bibitem{williams2015data}
M. O. Williams, I. G. Kevrekidis, and C. W. Rowley, ``A data--driven approximation of the koopman operator: Extending dynamic mode decomposition,'' \textit{J. Nonlinear Sci.}, vol. 25, pp.1307--1346, Jun. 2015.

\bibitem{baddoo2023physics}
P. J. Baddoo, B. Herrmann, B. J. McKeon, J. Nathan Kutz, and S. L. Brunton, ``Physics-informed dynamic mode decomposition,'' \textit{Proc. Royal Soc. A}, vol. 479, p. 20220576, Mar. 2023.

\bibitem{takeishi2017subspace}
N. Takeishi, Y. Kawahara, and T. Yairi, ``Subspace dynamic mode decomposition for stochastic Koopman analysis,'' \textit{Phys. Rev. E}, vol. 96, no. 3, p. 033310, Jun. 2017.

\bibitem{dmdspectrum}
B. W.Brunton, L. A. Johnson, J. G. Ojemann, and J. N. Kutz, ``Extracting spatial-temporal coherent patterns in large-scale neural recordings using dynamic mode decomposition,'' \textit{J. Neurosci. Methods}, vol. 258, pp. 1--15, Jan. 2016.

\bibitem{vecchiato2019electroencephalographic}
G. Vecchiato \textit{et al.}, ``Electroencephalographic time-frequency patterns of braking and acceleration movement preparation in car driving simulation,'' \textit{Brain Res.}, vol. 1716, pp. 16--26, Aug. 2019.

\bibitem{welch1967use}
P. Welch, ``The use of fast fourier transform for the estimation of power spectra$\colon$ A method based on time averaging over short, modified periodograms,'' \textit{IEEE Trans. Audio Electroacoust.}, vol. 15, no. 2, pp. 70--73, Jun. 1967.

\end{thebibliography}

\begin{IEEEbiography}[{\includegraphics[width=1in,height=1.25in,clip,keepaspectratio]{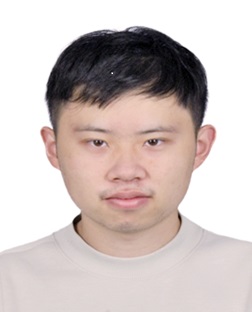}}]{Zikun Zhou} received B.S. degree in Vehicle Engineering from Beijing Institute of Technology, Beijing, China, in 2023, where he is currently working toward the Ph.D. degree in vehicle engineering. His research interests include brain–computer interface, human–vehicle interaction, and driving intention prediction.
\end{IEEEbiography}

\begin{IEEEbiography}[{\includegraphics[width=1in,height=1.25in,clip,keepaspectratio]{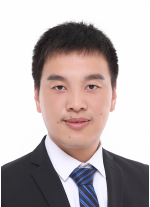}}]{Wenshuo Wang} (SM'15-M'18) received his Ph.D. degree in mechanical engineering from the Beijing Institute of Technology (BIT) in 2018.  Presently, he is a Full Professor at the School of Mechanical Engineering, BIT, Beijing, China. Prior to his role at BIT, he completed Postdoctoral fellowships at McGill University, Carnegie Mellon University (CMU), and UC Berkeley between 2018 and 2023. Furthermore, from 2015 to 2018, he served as a Research Assistant at UC Berkeley and the University of Michigan, Ann Arbor. His research interests focus on Bayesian nonparametric learning, human driver model, human–vehicle interaction, ADAS, and autonomous vehicles.
\end{IEEEbiography}

\begin{IEEEbiography}[{\includegraphics[width=1in,height=1.25in,clip,keepaspectratio]{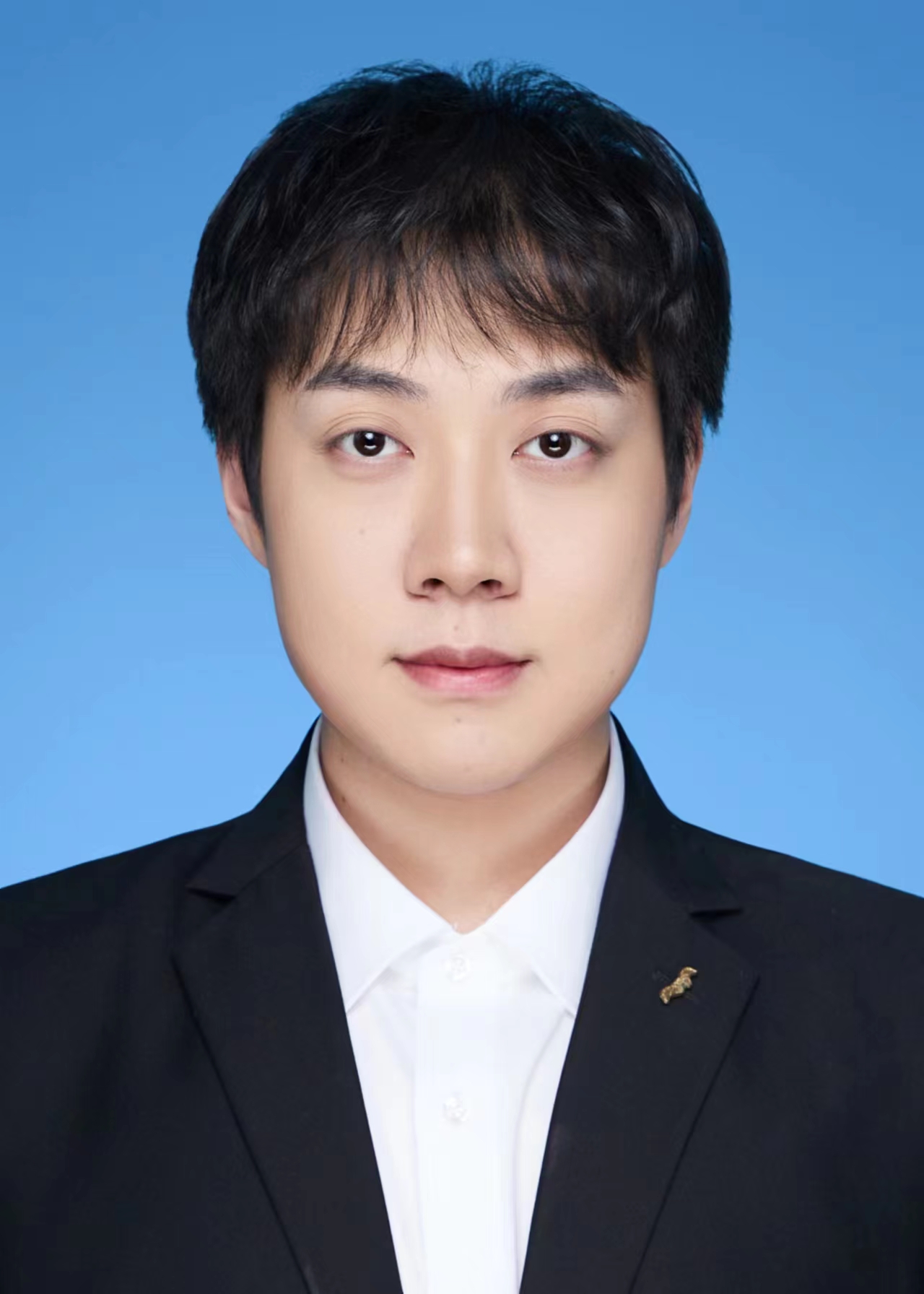}}]{Wenzhuo Liu (Graduate Student Member, IEEE)} was born in Jinan, Shandong Province, China in 1999. He received his master's degree in computer science and technology from China University of Mining and Technology (Beijing). He was a joint student at the State Key Laboratory of Intelligent Green Vehicles and Mobility, School of Vehicle and Mobility, Tsinghua University for three years. Currently, he is a PhD candidate at the State Key Laboratory of Intelligent Unmanned Systems Technology, Beijing Institute of Technology (BIT), China. His research interests include multi-task learning, autonomous driving perception, multi-modal fusion, and driver intention recognition.
\end{IEEEbiography}

\begin{IEEEbiography}[{\includegraphics[width=1in,height=1.25in,clip,keepaspectratio]{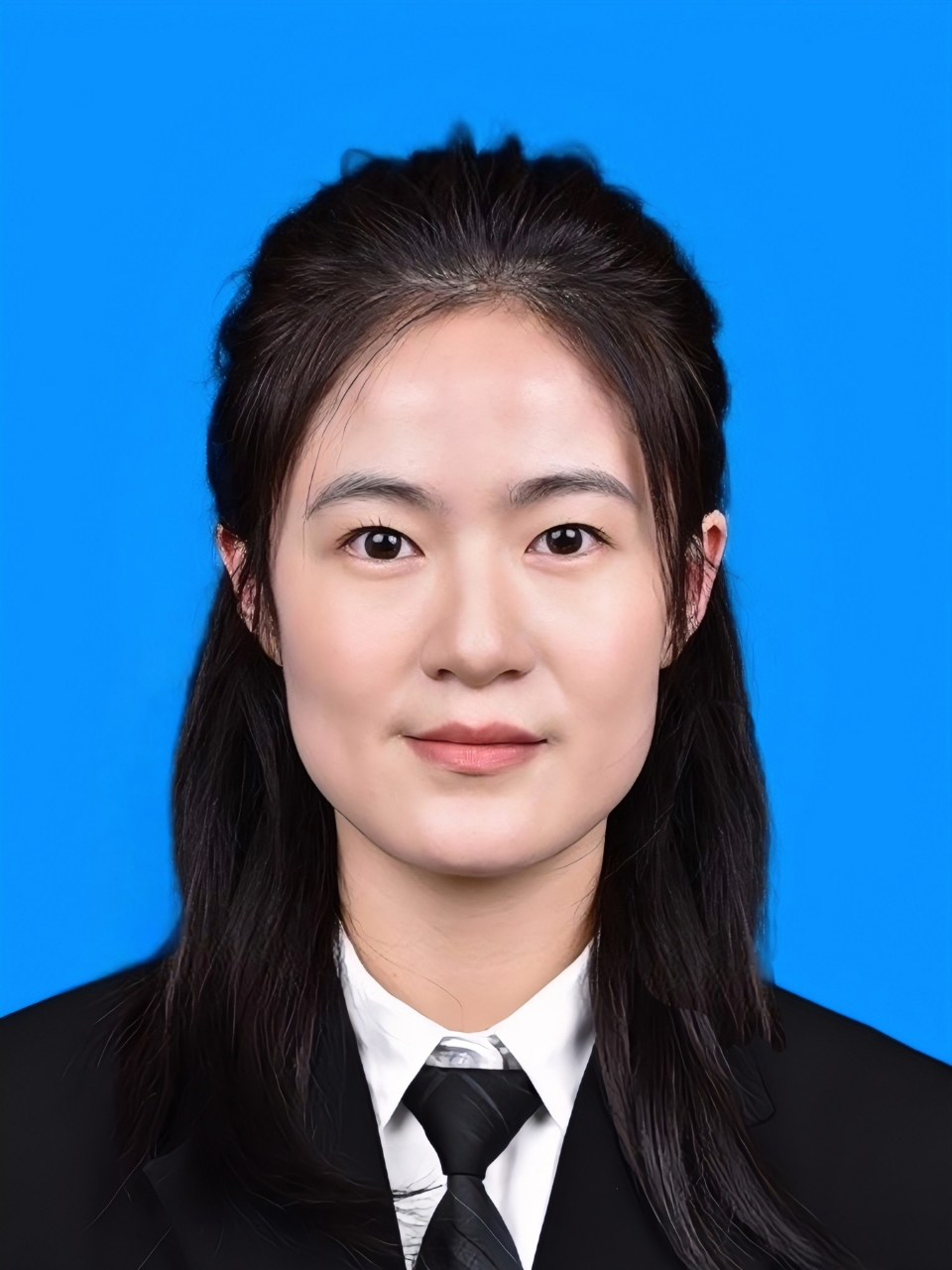}}]{Hui Yao} received the B.S. degree in mechanical engineering from Beijing Institute of Technology (BIT), Beijing, China, in 2018. She is currently working toward the Ph.D. degree in mechanical engineering from Beijing Institute of Technology (BIT), Beijing, China.
Her research interests include driver intention detection, machine learning, brain-machine intelligence and human-intelligence vehicle collaboration.
\end{IEEEbiography}

\begin{IEEEbiography}[{\includegraphics[width=1in,height=1.25in,clip,keepaspectratio]{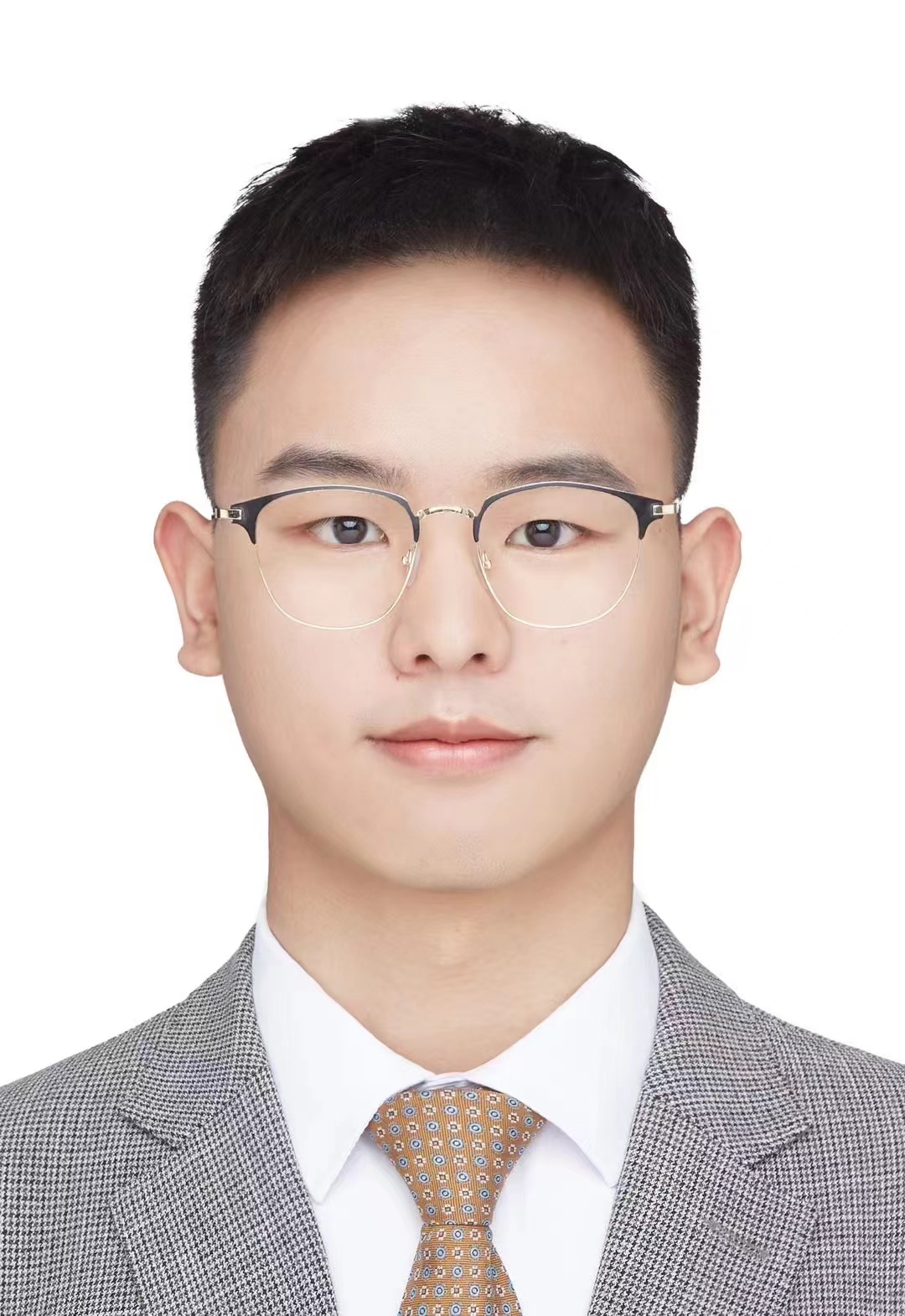}}]{Chaopeng Zhang} received B.S. degree in Mechanical Engineering from Beijing Institute of Technology, Beijing, China, in 2019, where he is currently working toward the Ph.D. degree in mechanical engineering.  His research interests include human factors in intelligent vehicles, human driver model, driving style recognition, and driving intention recognition.
\end{IEEEbiography}

\begin{IEEEbiography}[{\includegraphics[width=1in,height=1.25in,clip,keepaspectratio]{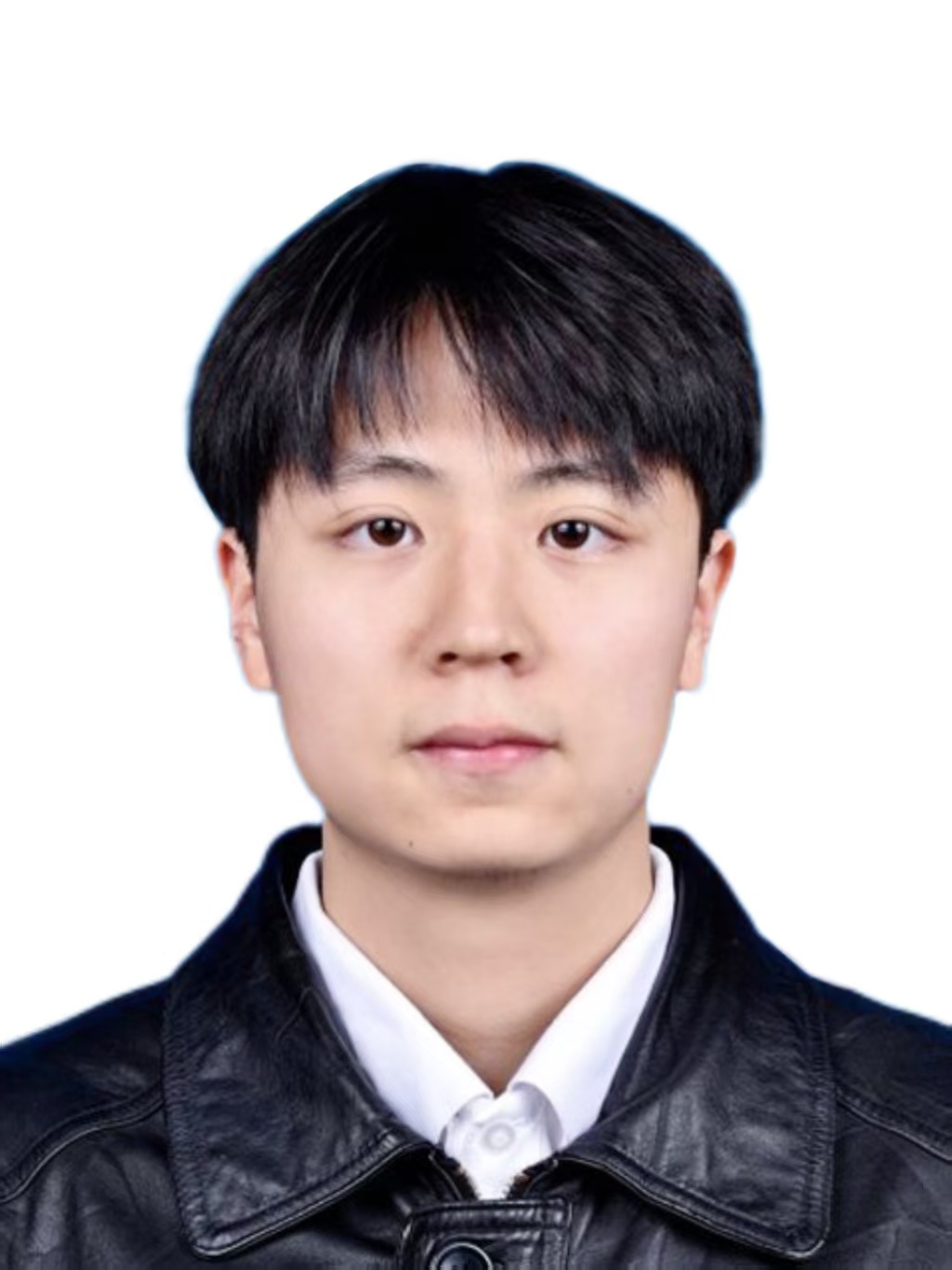}}]{Yichen Liu} received B.S. degree in Vehicle Engineering from Beijing Institute of Technology, China, in 2024, where he is currently pursuing the Master's degree in Vehicle engineering. His research interests include intelligent vehicles, human-vehicle interaction, and subjective-objective consistency evaluation.
\end{IEEEbiography}

\begin{IEEEbiography}
[{\includegraphics[width=1in,height=1.25in,clip,keepaspectratio]{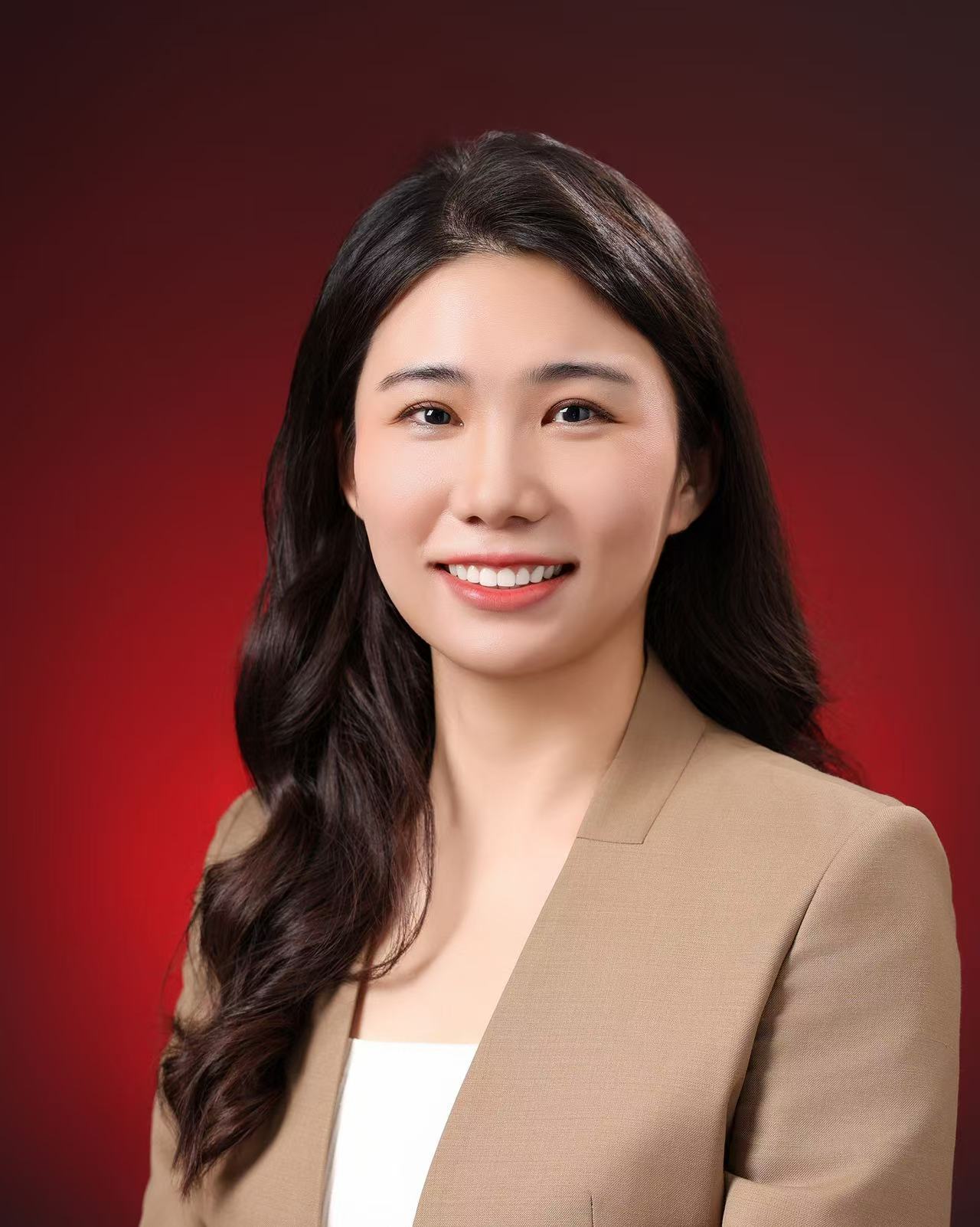}}]{Xiaonan Yang} is an Associate Professor at Beijing Institute of Technology. Her research focus on human-centered AI (HCAI), emphasizes human cognitive and behavioral characteristics in human-in-the-loop systems, encompassing natural human-computer interaction, multimodal intention recognition, cognitive load, visual cognitive mechanisms, and visual fatigue.
\end{IEEEbiography}

\begin{IEEEbiography}[{\includegraphics[width=1in,height=1.25in,clip,keepaspectratio]{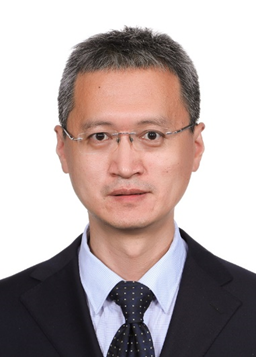}}]{Junqiang Xi} received the B.S. degree in automotive engineering from the Harbin Institute of Technology, Harbin, China, in 1995, and the Ph.D. degree in vehicle engineering from the Beijing Institute of Technology (BIT), Beijing, China, in 2001. In 2001, he joined the State Key Laboratory of Vehicle Transmission, BIT. During 2012–2013, he made research as an Advanced Research Scholar in Vehicle Dynamic and Control Laboratory, Ohio State University, Columbus, OH, USA. He is currently a Professor and Director of Automotive Research Center in BIT. His research interests include vehicle dynamic and control, powertrain control, mechanics, intelligent transportation system, and intelligent vehicles.
\end{IEEEbiography}

\end{document}